\documentclass[prb,twocolumn,floatfix,superscriptaddress,twoside]{revtex4}

\usepackage{dcolumn}
\usepackage{graphicx,subfigure}
\usepackage{amsmath,amssymb}

\newcommand{\nx}{n_{\textrm{X}}}

\newcommand{\cf}{\textit{cf.}{ }}
\newcommand{\etal}{\textit{et al.}{ }}
\newcommand{\ie}{\textit{i.e.}{ }}

\newcolumntype{.}[1]{D{.}{.}{#1}}

\begin{document}

\title{Nucleation of Al$_3$Zr and Al$_3$Sc in aluminum alloys: \\
from kinetic Monte Carlo simulations to classical theory.}

\author{Emmanuel \surname{Clouet}}
\email{emmanuel.clouet@cea.fr}
\affiliation{Pechiney Centre de Recherches de Voreppe, B.P.~27,
38341  Voreppe~cedex, France}
\affiliation{Service de Recherches de M\'etallurgie Physique, CEA/Saclay,
91191 Gif-sur-Yvette, France}
\author{Maylise \surname{Nastar}}
\affiliation{Service de Recherches de M\'etallurgie Physique, CEA/Saclay,
91191 Gif-sur-Yvette, France}
\author{Christophe \surname{Sigli}}
\affiliation{Pechiney Centre de Recherches de Voreppe, B.P.~27,
38341  Voreppe~cedex, France}


\date{\today}

\begin{abstract}
Zr and Sc precipitate in aluminum alloys to form the compounds
Al$_3$Zr and Al$_3$Sc which for low supersaturations of the solid
solution have the L1$_2$ structure. The aim of the present study
is to model at an atomic scale this kinetics of precipitation and
to build a mesoscopic model based on classical nucleation theory
so as to extend the field of supersaturations and annealing times
that can be simulated. We use some ab-initio calculations and
experimental data to fit an Ising model describing thermodynamics
of the Al-Zr and Al-Sc systems. Kinetic behavior is described by
means of an atom-vacancy exchange mechanism. This allows us to
simulate with a kinetic Monte Carlo algorithm kinetics of
precipitation of Al$_3$Zr and Al$_3$Sc. These kinetics are then
used to test the classical nucleation theory. In this purpose, we
deduce from our atomic model an isotropic interface free energy
which is consistent with the one deduced from experimental
kinetics and a nucleation free energy. We test different
mean-field approximations (Bragg-Williams approximation as well as
Cluster Variation Method) for these parameters. The classical
nucleation theory is coherent with the kinetic Monte Carlo
simulations only when CVM is used: it manages to reproduce the
cluster size distribution in the metastable solid solution and its
evolution as well as the steady-state nucleation rate. We also
find that the capillary approximation used in the classical
nucleation theory works surprisingly well when compared to a
direct calculation of the free energy of formation for small
L1$_2$ clusters.
\end{abstract}

\maketitle
\section{Introduction}

Precipitation kinetics of a metastable solid solution is known to
be divided in three successive stages: the nucleation, growth, and
coarsening of nuclei of the new stable phase. The first stage of
precipitation is of great practical interest but difficult to
observe experimentally. Kinetic Monte Carlo simulation is the
suitable tool for a numerical prediction of a nucleation kinetics
\cite{SOI02,BEL02} but a rationalization of the results is
difficult and atomic simulations cannot reach very low
supersaturations. On the other hand, classical descriptions of
these different stages \cite{MAR78,WAG91} are well established and
the associated models are now widely used to understand
experimental kinetics and to model technological processes
\cite{DES99,ROB01,STO02,MAU01}. Recently, classical nucleation
theory has been shown to be in good agreement with more reliable
atomic models by way of a direct comparison with kinetic Monte
Carlo simulations \cite{RAM99,NOV00,SHN99,SOI00,BER03}. These
different studies included decomposition of a metastable solid
solution for a demixing binary system  on a surface \cite{RAM99}
or in the bulk \cite{SHN99,SOI00} and kinetics of
electrodeposition on a surface \cite{BER03}. In this last study,
Berthier \etal show that physical parameters of classical
nucleation theory have to be carefully calculated so as to
reproduce atomic simulations. In the present paper, we want to
extend the range of comparison between classical nucleation theory
and atomic simulations by studying the case of an ordering system
on a frustrated lattice. We thus choose to model kinetics of
precipitation of a L1$_2$ ordered compound formed from a solid
solution lying on a face centered cubic (fcc) lattice.

For fcc lattice it is now well established that one has to use a
mean field approximation more accurate than the widely used
Bragg-Williams one in order to calculate thermodynamic properties
\cite{DEF94}. The Cluster Variation Method (CVM)
\cite{KIK51,SAN78} enables one to obtain phase diagrams which are
in quantitative agreement with thermodynamic Monte Carlo
simulations \cite{MOH85,FIN94}. When CVM is used, frustration
effects on the tetrahedron of first nearest neighbors and short
range order due to interactions are considered in a satisfying way
enabling one to predict quantitatively thermodynamic behavior.
Nevertheless, the use of CVM is often restricted to the
calculation of equilibrium properties and, thus, for computing
thermodynamic properties of the metastable supersaturated solid
solution in classical nucleation theory one merely considers
Bragg-Williams approximation. The purpose of this paper is then to
show that the use of CVM calculations with classical nucleation
theory leads to a satisfying description of the metastable solid
solution and extend the range of supersaturations that can be
modelled with this theory.

In this purpose we build an atomic model which allows us to study
kinetics of precipitation of Al$_3$Zr and Al$_3$Sc. The two
considered binary systems, Al-Zr and Al-Sc, have different kinetic
properties: the interaction with vacancies is repulsive for Zr
atoms whereas it is attractive for Sc atoms. On the other hand,
for low supersaturations, thermodynamics of both systems are quite
similar. Al$_3$Zr has the stable DO$_{23}$
structure\cite{PEARSON}, but for small supersaturations of the
solid solution, Al$_3$Zr precipitates with the metastable L1$_2$
structure and
precipitates with the DO$_{23}$ structure only appear for
prolonged heat treatment and high enough supersaturations
\cite{RYU69,ROB01,NES72}. As for Al$_3$Sc, the stable structure is
L1$_2$ \cite{PEARSON} and thus only L1$_2$ precipitates have been
observed during experimental kinetics \cite{HYL92,MAR01,NOV01}. In
this study we mainly focus on the nucleation stage and therefore
we consider that both Zr and Sc lead to the precipitation of a
compound having the L1$_2$ structure. In this context, Al-Zr and
Al-Sc systems are really similar from a thermodynamic point of
view unlike their kinetic behavior. It is then interesting to
study these two systems in parallel and to see if classical
nucleation theory manages to reproduce atomic simulations for
these two different kinetic behaviors.

The atomic model used in kinetic Monte Carlo simulations is built
using experimental data as well as ab-initio calculations. We
deduce from it physical parameters entering mesoscopic models like
classical nucleation theory and show how this theory compares to
atomic simulations for different supersaturations and different
annealing temperatures. The capillary approximation used in
classical nucleation theory is then discussed as well as different
mean field approximations that can be combined
with it.

\section{Atomic model}

\subsection{Al-Zr and Al-Sc thermodynamics}

In order to simulate thermodynamic behavior of Al-Zr and Al-Sc
binary systems, we use a rigid lattice: configurations of the
system are described by the occupation numbers $p^{i}_{n}$ with
$p^{i}_{n}=1$ if the site $n$ is occupied by an atom of type $i$
and $p^{i}_{n}=0$ if not. Energies of such configurations are
given by an Ising model with first and second nearest neighbor
interactions. This is the simplest model to simulate precipitation
of a stoichiometric Al$_3$X compound in the L1$_2$ structure.
Indeed one has to include second nearest neighbor interactions,
otherwise L1$_2$ precipitates do not show perfect Al$_3$X
composition. On the other hand, there is no use to consider
interactions beyond second nearest neighbors as these interactions
are significantly lower than first and second nearest neighbor
interactions\cite{CLO02}. We could have considered interactions
for clusters other than pairs too, but we showed that the use of
interactions for first nearest neighbor triangle and tetrahedron
does not really change the kinetics of precipitation\cite{CLO02b}:
the Onsager coefficients defining diffusion in the solid solution
are unchanged with or without these interactions as well as the
nucleation free energy. Thus, in our model, the energy per site of
a given configuration is
\begin{equation}
E = \frac{1}{2 N_s} \sum_{\substack{ n,m \\ i,j}}
\epsilon_{ij}^{(1)} p^{i}_{n} p^{j}_{m} + \frac{1}{2 N_s}
\sum_{\substack{ r,s \\ i,j}} \epsilon_{ij}^{(2)} p^{i}_{r}
p^{j}_{s} , \label{Ising}
\end{equation}
where the first and second sums respectively runs on all first and
second nearest neighbor pairs of sites, $N_s$ is the number of
lattice sites, $\epsilon_{ij}^{(1)}$ and $\epsilon_{ij}^{(2)}$ are
the respective effective energies of a first and second nearest
neighbor pair in the configuration \{$i,j$\}.

\begin{table}
\caption{First and second nearest neighbor pair effective energies
(in eV). Only interactions different from zero are presented.}
\label{energies_tab}
\begin{ruledtabular}
\begin{tabular}{lclclcl}
$\epsilon_{\textrm{AlAl}}^{(1)}$ &=& $-0.560$ &\hspace*{15ex}&
$\epsilon_{\textrm{AlV}}^{(1)}$ &=& $-0.222$ \\
$\epsilon_{\textrm{ZrZr}}^{(1)}$ &=& $-1.045$ &\hspace*{15ex}&
$\epsilon_{\textrm{ZrV}}^{(1)}$ &=& $-0.350$ \\
$\epsilon_{\textrm{ScSc}}^{(1)}$ &=& $-0.650$ &\hspace*{15ex}&
$\epsilon_{\textrm{ScV}}^{(1)}$ &=& $-0.757$ \\
$\epsilon_{\textrm{AlZr}}^{(1)}$ &=&
\multicolumn{5}{l}{$-0.979+24.4\times10^{-6}T$} \\
$\epsilon_{\textrm{AlSc}}^{(1)}$ &=&
\multicolumn{5}{l}{$-0.759+21.0\times10^{-6}T$} \\
$\epsilon_{\textrm{VV}}^{(1)}$ &=& $-0.084$\\
$\epsilon_{\textrm{AlZr}}^{(2)}$ &=&
\multicolumn{5}{l}{$+0.101-22.3\times10^{-6}T$} \\
$\epsilon_{\textrm{AlSc}}^{(2)}$ &=&
\multicolumn{5}{l}{$+0.113-33.4\times10^{-6}T$} \\
\end{tabular}
\end{ruledtabular}
\end{table}

With such a model, as long as vacancy concentration can be
neglected, thermodynamic behavior of Al-X system (X~$\equiv$ Zr or
Sc) only depends on the order energies
\begin{eqnarray}
\omega^{(1)} &=& \epsilon_{\textrm{AlX}}^{(1)}
-\frac{1}{2}\epsilon_{\textrm{AlAl}}^{(1)}
-\frac{1}{2}\epsilon_{\textrm{XX}}^{(1)} , \\
\omega^{(2)} &=& \epsilon_{\textrm{AlX}}^{(2)}
-\frac{1}{2}\epsilon_{\textrm{AlAl}}^{(2)}
-\frac{1}{2}\epsilon_{\textrm{XX}}^{(2)} .
\end{eqnarray}
We use experimental data combined with ab-initio calculations to
obtain these order energies for Al-Zr and Al-Sc systems.

First nearest neighbor order energies are chosen so as to
correctly reproduce formation energies of Al$_3$Zr and Al$_3$Sc
compounds in L1$_2$ structure, $\Delta F(\textrm{Al}_3\textrm{X,
L1}_2) = 3\omega^{(1)}$. For Al$_3$Zr, we use the free energy of
formation that we previously calculated\cite{CLO02}. For Al$_3$Sc,
we calculate the enthalpy of formation with the full-potential
linear-muffin-tin-orbital method\cite{MET93} in the generalized
gradient approximation\cite{PER96} and we use the value of
Ref.~\onlinecite{AST01} and \onlinecite{OZO01} for the vibrational
contribution to the free energy of formation:
\begin{eqnarray}
\Delta F(\textrm{Al}_3\textrm{Zr, L1}_2)&=& -0.530 +
73.2\times10^{-6}T
\textrm{~eV} \nonumber , \\
\Delta F(\textrm{Al}_3\textrm{Sc, L1}_2)&=& -0.463 +
62.9\times10^{-6}T \textrm{~eV} \nonumber .
\end{eqnarray}

Second nearest neighbor interactions are chosen so as to reproduce
Zr and Sc solubility limits in Al. Indeed these limits only depend
on order energy $\omega^{(2)}$, as can be seen from the low
temperature expansion\cite{DUCASTELLE} to the second order in
excitation energies:
\begin{equation}
x_{\textrm{X}}^{eq} = \exp{\left(-6\omega^{(2)}/kT\right)} + 6
\exp{\left(-10\omega^{(2)}/kT\right)} .
\end{equation}
We check using the CVM in the tetrahedron-octahedron
approximation\cite{KIK51,SAN78} that this low temperature
expansion for the solubility limit is correct in the whole range
of temperature of interest, \ie until Al melting temperature
($T^{mel}=934$K). For Al-Zr interactions, as we want to model
precipitation of the metastable L1$_2$ structure of Al$_3$Zr
compound, we use the metastable solubility limit that we
previously obtained from ab-initio calculations\cite{CLO02},
whereas for Al$_3$Sc the L1$_2$ structure is stable and we use the
solubility limit arising from a thermodynamic modelling of
experimental data \cite{MUR98}:
\begin{eqnarray}
x_{\textrm{Zr}}^{eq} &=& \exp{ \left( (-0.620 + 155\times10^{-6}T)
\textrm{~eV} / kT \right)} \nonumber , \\
x_{\textrm{Sc}}^{eq} &=& \exp{ \left( (-0.701 + 230\times10^{-6}T)
\textrm{~eV} / kT \right)} \nonumber .
\end{eqnarray}
One should notice that these solubility limits have been found to
be consistent with ab-initio calculations \cite{CLO02,OZO01} and
thus with the formation energies we used for Al$_3$Zr and and
Al$_3$Sc.

Unlike thermodynamics, kinetics does not only depend on order
energies but also on effective energies $\epsilon^{(1)}_{ij}$ and
$\epsilon^{(2)}_{ij}$. We deduce them from $\omega^{(1)}$ and
$\omega^{(2)}$ by using experimental values for cohesive energies
of pure elements\cite{HULTGREN}: $E^{coh}(\textrm{Al}) = 3.36$~eV,
$E^{coh}(\textrm{Zr}) =6.27$~eV, and $E^{coh}(\textrm{Sc})
=3.90$~eV. We assume that second nearest neighbor interactions do
not contribute to these cohesive energies, \ie
$\epsilon^{(2)}_{\textrm{XX}}=0$ (X~$\equiv$ Al, Zr, or Sc) and we
neglect any possible temperature dependence of these energies.
Therefore, the cohesive energy of X element is
$E^{coh}(\textrm{X}) = 6\epsilon^{(1)}_{\textrm{XX}}$. Resulting
effective energies are presented in table~\ref{energies_tab}.

\subsection{Al-Zr and Al-Sc kinetics}

We introduce in the Ising model atom-vacancy interactions for
first nearest neighbors (Tab.~\ref{energies_tab}), so as to
consider the electronic relaxations around the vacancy. Without
these interactions, the vacancy formation energy
$E^{for}_{\textrm{V}}$ in a pure metal would necessarily equal the
cohesive energy which is in contradiction with experimental data.
$\epsilon_{\textrm{AlV}}^{(1)}$ and
$\epsilon_{\textrm{ZrV}}^{(1)}$ are deduced from vacancy formation
energy respectively in pure Al \cite{LANDOLT25},
$E^{for}_{\textrm{V}}=0.69$~eV, and in pure Zr \cite{LEB99},
$E^{for}_{\textrm{V}} = 2.07$~eV. For Zr, this energy corresponds
to the hcp structure which is quite similar to the fcc one (same
first nearest neighborhood). Therefore, we assume that the vacancy
energy is the same in both structures. This is then possible to
correct this formation energy to take into account the difference
between Al and Zr equilibrium volumes, but this leads to a
correction of $\sim10$\% for $E^{for}_{\textrm{V}}$ and does not
really change the physical interaction between Zr atoms and
vacancies. We thus choose to neglect such a correction. To compute
the interaction between Sc atoms and vacancies, we can directly
deduce $\epsilon_{\textrm{ScV}}^{(1)}$ from the experimental
binding energy in aluminum \cite{MIU00}, $E^{bin}_{\textrm{ScV}} =
\epsilon_{\textrm{AlV}}^{(1)} + \epsilon_{\textrm{AlSc}}^{(1)} -
\epsilon_{\textrm{ScV}}^{(1)} - \epsilon_{\textrm{AlAl}}^{(1)} =
0.35$~eV at 650~K. Such an experimental data does not exist for Zr
impurity, but we can check that the physical interaction we obtain
is correct. The binding energy deduced from our set of parameters
is strongly negative ($E^{bin}_{\textrm{ZrV}} = -0.276$~eV at
650~K). This is in agreement with the experimental fact that no
attraction has been observed between vacancy and Zr impurity
\cite{LANDOLT25,SIM77}. This repulsion in the case of Zr impurity
and this attraction in the case of Sc impurity are related to the
difference of cohesive energies between Zr and Sc, showing thus
that elastic relaxations around the vacancy are not the dominant
effect. It could explain why Zr diffusion coefficient in aluminum
is so low compared to the Sc one.
Some ab-initio calculations have been made to compute this binding
energy with a vacancy for all transition metals in aluminum
\cite{HOS96}. They obtained in the case of Zr as well as Sc
impurity a repulsive interaction with a vacancy. This is in
contradiction with the experimental data we use to compute
Sc-vacancy interaction. Such a disagreement may arise from
approximations made in the calculation (Kohn-Korringa-Rostoker
Green's function method) as the neglect of atom relaxations and
the box that includes only the first nearest neighbors of the
impurity-vacancy complex. Nevertheless these ab-initio
calculations showed that a binding energy as large as 0.35~eV is
possible as the value obtained for Sr impurity was even larger.

We use the experimental value of the divacancy binding energy
\cite{LANDOLT25}, $E^{bin}_{2\textrm{V}}=0.2$~eV, in order to
compute a vacancy-vacancy interaction,
$\epsilon_{\textrm{VV}}^{(1)} = 2\epsilon_{\textrm{AlV}}^{(1)} -
\epsilon_{\textrm{AlAl}}^{(1)} -  E^{bin}_{2\textrm{V}}$. If we do
not include this interaction and set it equal to zero instead, we
obtain a binding energy which is slightly too low, divacancies
being thus not as stable as they should be.
Some recent ab-initio calculations \cite{CAR00,CAR03} have shown
that divacancies should be actually unstable, the non-Arrhenius
temperature dependence of the vacancy concentration arising from
anharmonic atomic vibrations. Nevertheless, this does not affect
our Monte Carlo simulations as we only include one vacancy in the
simulation box, but this divacancy binding energy should be
considered more seriously if one wants to build a mean field
approximation of our diffusion model or if one wants to compensate
vacancy trapping by adding new vacancies in the simulation box.

\begin{table}
\caption{Kinetic parameters: contribution of the jumping atom to
the saddle point energy $e^{sp}_{\alpha}$ and attempt frequency
$\nu_{\alpha}$ for $\alpha$~$\equiv$ Al, Zr, and Sc atoms.}
\label{cin_tab}
\begin{ruledtabular}
\begin{tabular}{ccrcccr}
$e^{sp}_{\textrm{Al}}$ &=& $-8.219$~eV  &\hspace*{15ex}&
$\nu_{\textrm{Al}}$ &=& $1.36\times 10^{14}$~Hz \\
$e^{sp}_{\textrm{Zr}}$ &=& $-11.464$~eV &\hspace{15ex}&
$\nu_{\textrm{Zr}}$ &=& $9\times 10^{16}$~Hz \\
$e^{sp}_{\textrm{Sc}}$ &=& $-9.434$~eV& \hspace{15ex}&
$\nu_{\textrm{Sc}}$ &=& $4\times 10^{15}$~Hz\\
\end{tabular}
\end{ruledtabular}
\end{table}

Diffusion is described through vacancy jumps. The vacancy exchange
frequency with one of its twelve first nearest neighbors of type
$\alpha$ is given by
\begin{equation}
\Gamma_{\alpha\textrm{-V}}=\nu_{\alpha}
\exp{\left(-\frac{E^{act}_{\alpha}}{kT} \right)} , \label{rate_eq}
\end{equation}
where $\nu_{\alpha}$ is an attempt frequency and the activation
energy $E^{act}_{\alpha}$ is the energy change required to move
the $\alpha$ atom from its initial stable position to the saddle
point position. It is computed as the difference between the
contribution $e^{sp}_{\alpha}$ of the jumping atom to the saddle
point energy and the contributions of the vacancy and of the
jumping atom to the initial energy of the stable position. This
last contribution is obtained by considering all bonds which are
broken by the jump.

The attempt frequency $\nu_{\alpha}$ and the contribution
$e^{sp}_{\alpha}$of the jumping atom to the saddle point energy
can depend on the configuration \cite{NAS97,VAN01,LEB02}.
Nevertheless, we do not have enough information to see if such a
dependence holds in the case of Al-Zr or Al-Sc alloys. We thus
assume that these parameters depend only on the nature of the
jumping atom. We fit the six resulting kinetic parameters
(Tab.~\ref{cin_tab}) so as to reproduce Al self-diffusion
coefficient\cite{LANDOLT26} and Zr\cite{LANDOLT26,MAR73} and
Sc\cite{FUJ97} impurity diffusion coefficients:
\begin{eqnarray}
D_{\textrm{Al}^*} &=& 0.173\times10^{-4}
\exp{(-1.30\textrm{~eV}/kT)}
\textrm{~m}^{2}.\textrm{s}^{-1} \nonumber , \\
D_{\textrm{Zr}^*} &=& 728\times10^{-4}
\exp{(-2.51\textrm{~eV}/kT)}
\textrm{~m}^{2}.\textrm{s}^{-1} \nonumber , \\
D_{\textrm{Sc}^*} &=& 5.31\times10^{-4}
\exp{(-1.79\textrm{~eV}/kT)} \textrm{~m}^{2}.\textrm{s}^{-1}
\nonumber .
\end{eqnarray}
Zr is diffusing slower than Sc which itself is diffusing slower
than Al. The difference between diffusion coefficients is
decreasing with the temperature, but at the maximal temperature we
considered, \ie $T=873$~K, it still remains important as we then
have $D_{\textrm{Al}^*} \sim 20 D_{\textrm{Sc}^*} \sim 2000
D_{\textrm{Zr}^*}$.

\subsection{Monte Carlo simulations}
\label{KMC_chap}

We use residence time algorithm to run kinetic Monte Carlo
simulations. The simulation boxes contain $N_s=100^3$ or $200^3$
lattice sites and a vacancy occupies one of these sites. At each
step, the vacancy can jump with one of its twelve first nearest
neighbors, the probability of each jump being given by
Eq.~(\ref{rate_eq}). The time increment corresponding to this
event is
\begin{equation}
\Delta t = \frac{1}{N_s(1-13x^0_{\textrm{X}})
C_{\textrm{V}}(\textrm{Al}) }
\frac{1}{\sum_{\alpha=1}^{12}{\Gamma_{\alpha\textrm{-V}}}},
\label{time_inc_eq}
\end{equation}
where $x^0_{\textrm{X}}$ is the nominal concentration of the
simulation box (X~$\equiv$ Zr or Sc) and
$C_{\textrm{V}}(\textrm{Al})$ the real vacancy concentration in
pure Al as deduced from energy parameters of table
\ref{energies_tab}. The first factor appearing in
Eq.(\ref{time_inc_eq}) is due to the difference between the
experimental vacancy concentration in pure Al and the one observed
during the simulations. The dependence of this factor with the
concentration $x^0_{\textrm{X}}$ reflects that for each impurity
the corresponding lattice site and its twelve first nearest
neighbors cannot be considered as being pure Al. It is correct
only for a random solid solution in the dilute limit, but
concentrations considered in this study are low enough so the same
expression can be kept for this factor. The absolute time scale is
then obtained by summing only configurations where the vacancy is
surrounded by Al atoms in its first nearest neighborhood, \ie
where the vacancy is in pure Al. This ensures that the influence
on the time scale of the thermodynamic interaction between the
vacancy and the Sc or Zr impurity is correctly taken into account.
The method employed here is equivalent to measuring the fraction
of time spent by the vacancy in pure Al during the simulation and
multiplying then all time increments by this factor as done in
Ref.~\onlinecite{LEB02}. We check by running Monte Carlo
simulations for different sizes of the box that results do not
depend on the effective vacancy concentration.

For low impurity concentration, residence time algorithm can be
sped up by noticing that in most of the explored configurations
the vacancy is located in pure Al, \ie on a lattice site where all
exchange frequencies are equal to the one in pure Al. In such a
configuration, the move of the vacancy can be associated to a
random walk and the corresponding time increment is known \emph{a
priori}. Lattice sites corresponding to such configurations can be
detected in the beginning of the simulation and the corresponding
tables need to be modified only each time the vacancy exchanges
with an impurity \cite{DAL02}. For impurity concentration in the
range $5\times10^{-3}\leq x^0_{\textrm{X}} \leq 1\times10^{-2}$,
the algorithm is sped up by a factor $\sim2$. This allows us to
simulate lower supersaturations of the solid solution than we
could have done with a conventional algorithm.

\begin{figure*}
\subfigure[$t=24$~ms]{
\includegraphics[width=0.235\textwidth]{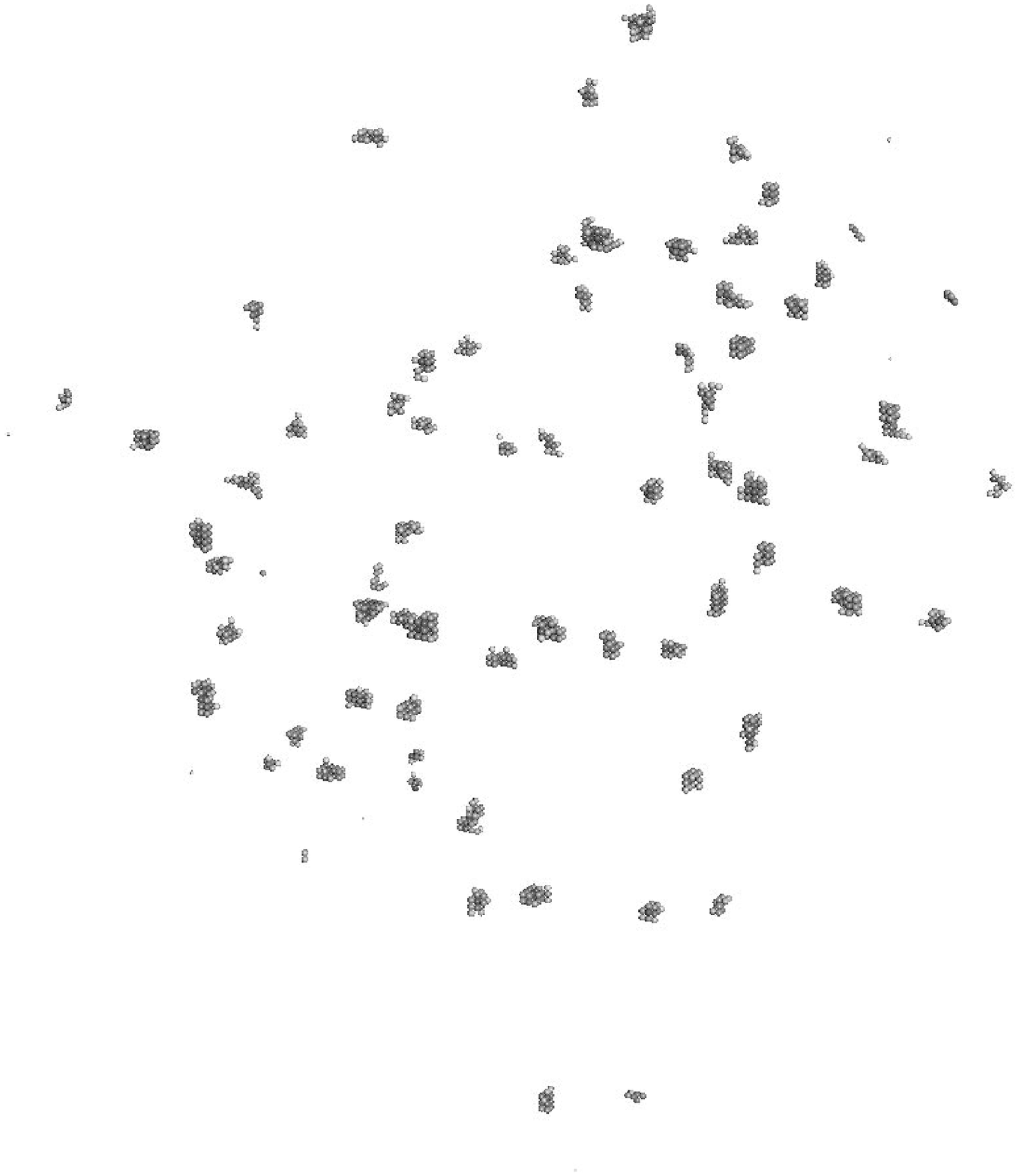}}
\subfigure[$t=60$~ms]{
\includegraphics[width=0.235\textwidth]{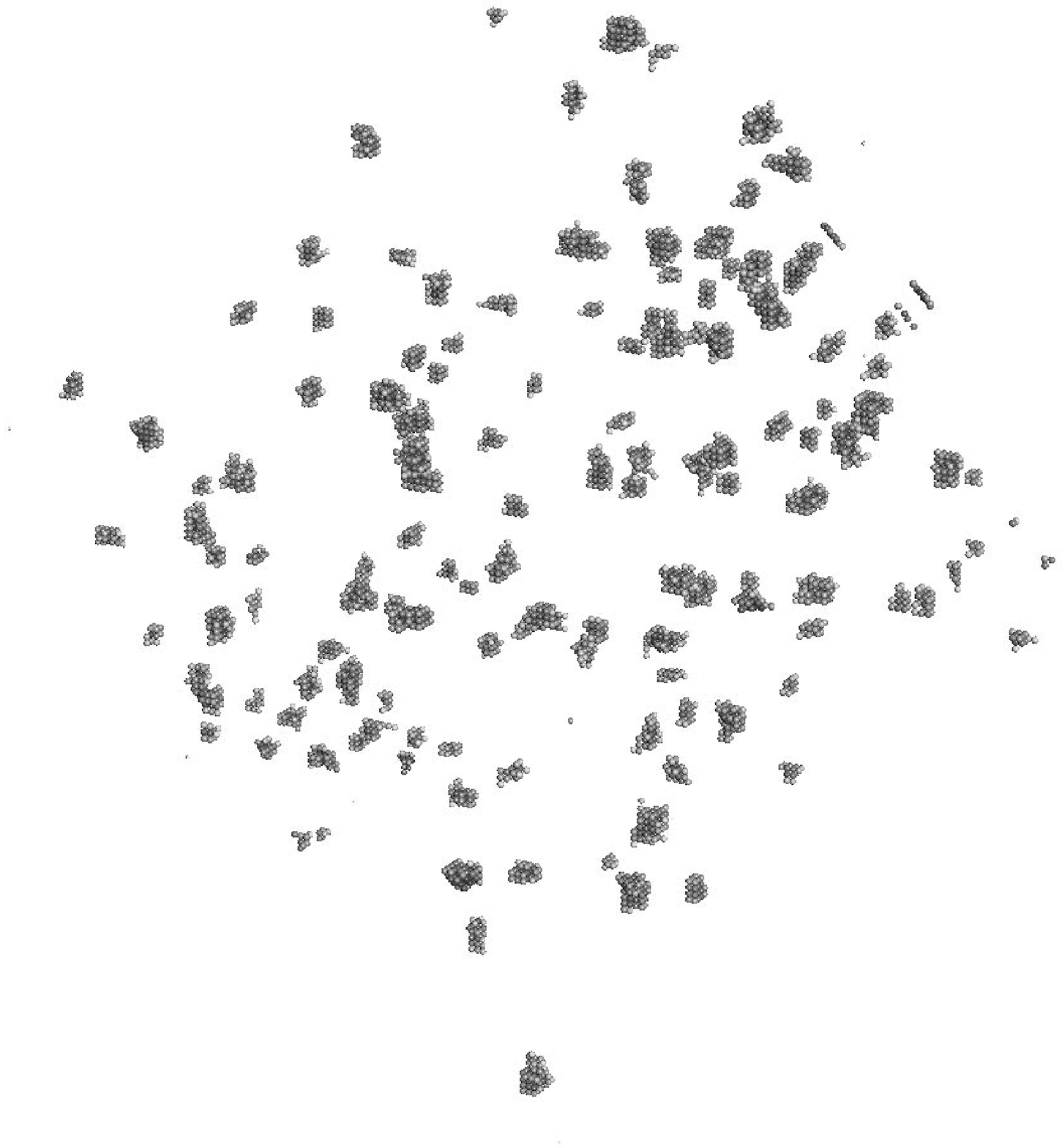}}
\subfigure[$t=151$~ms]{
\includegraphics[width=0.235\textwidth]{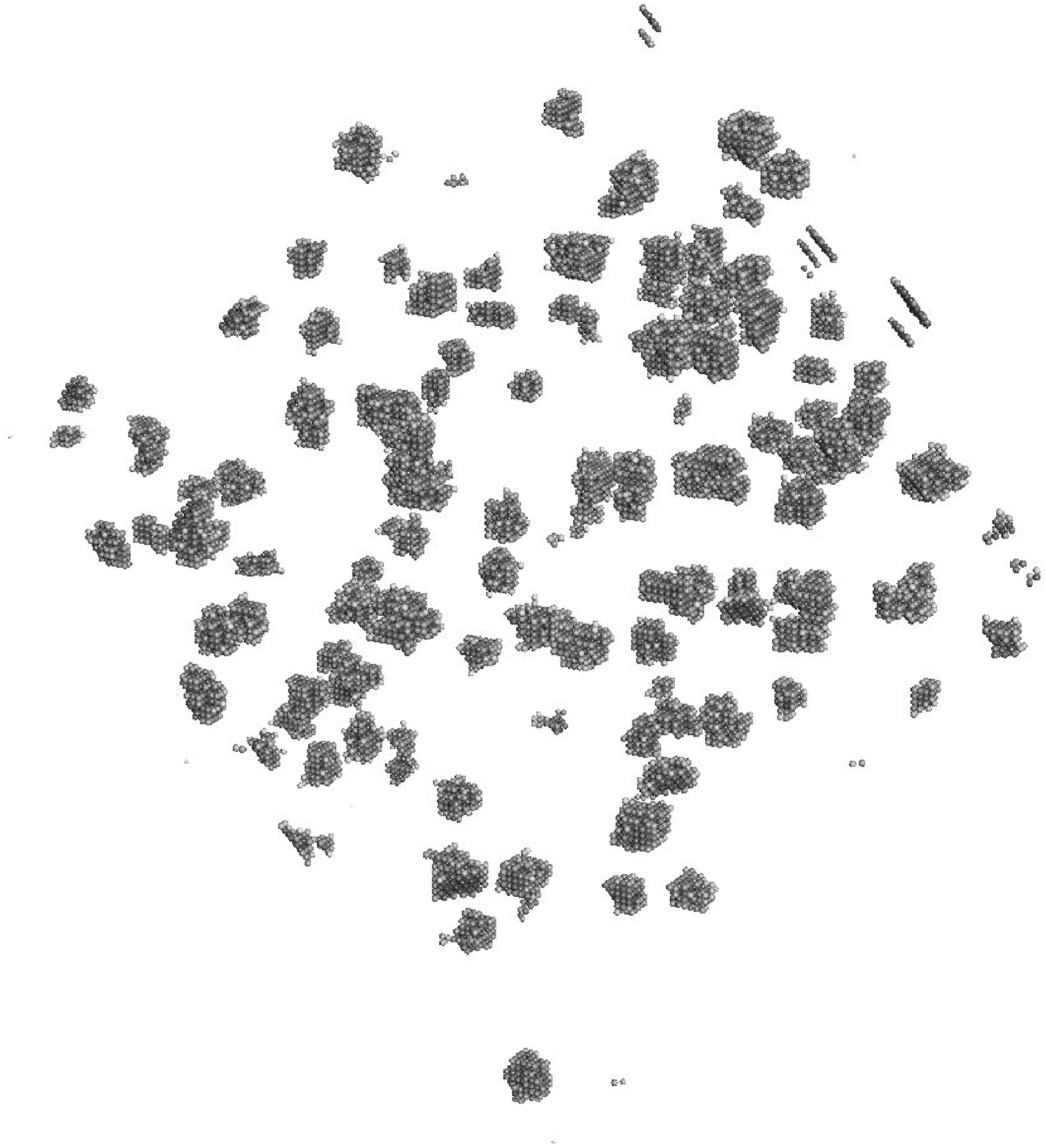}}
\subfigure[$t=366$~ms]{
\includegraphics[width=0.235\textwidth]{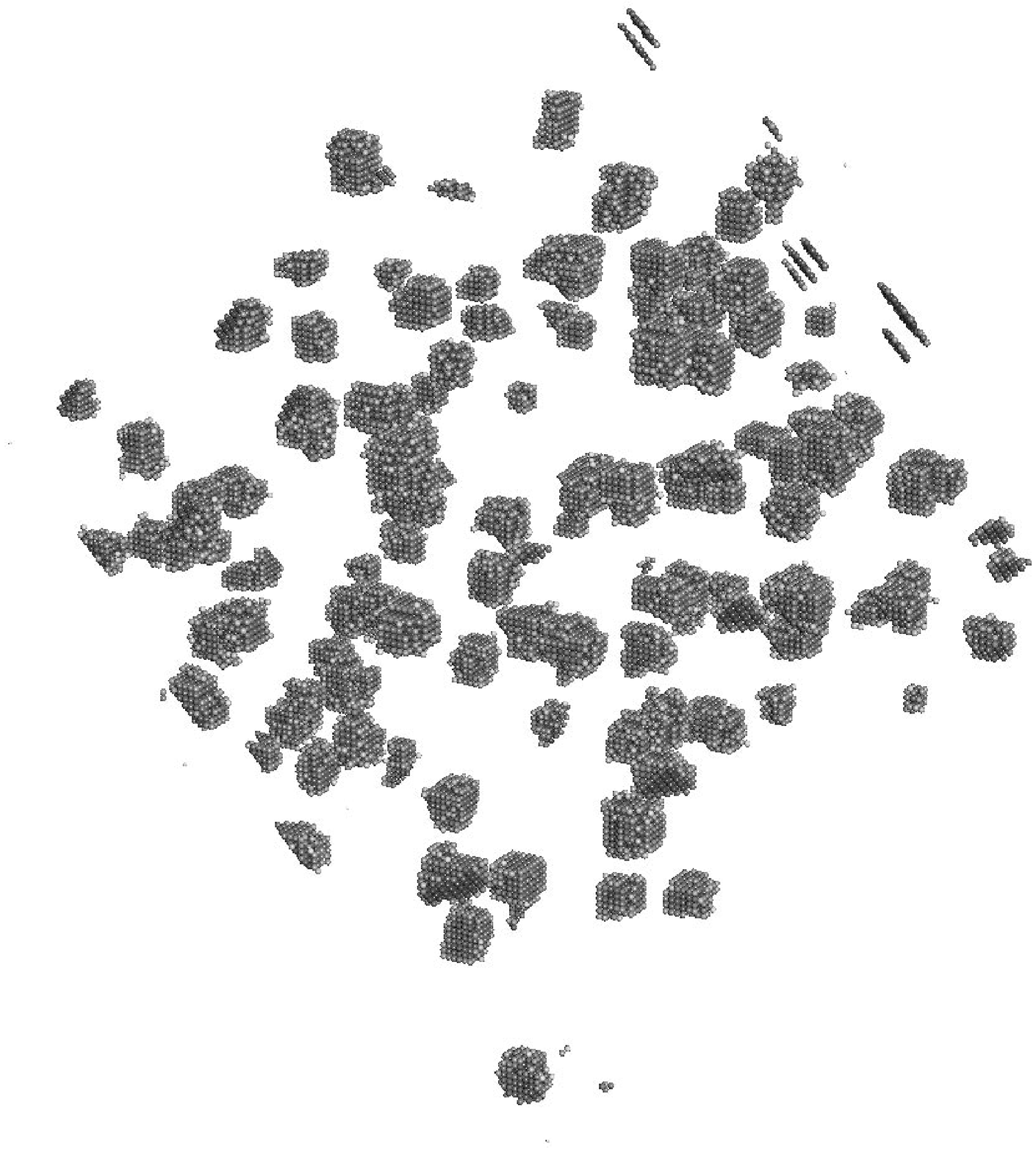}}
\caption{Monte Carlo simulation of the kinetics of precipitation
of Al$_3$Sc for a supersaturated aluminum solid solution of
nominal concentration $x^0_{\textrm{Sc}}=0.005$ at $T=773$~K. The
simulation box contains $8\times10^6$ lattice sites. Only Sc atoms
belonging to L1$_2$ precipitates are shown. The critical size used
is $n^*_{\textrm{Sc}}=13$.} \label{alsc_kinetic_picture}
\end{figure*}

\begin{figure}
\includegraphics[width=0.9\linewidth]{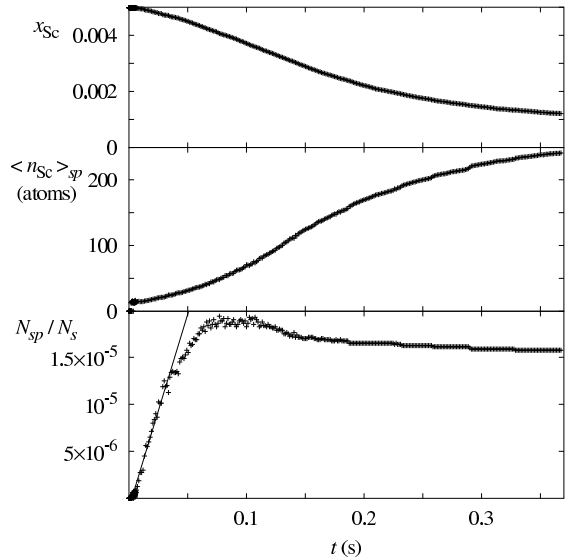}
\caption{Kinetics of precipitation of a supersaturated aluminum
solid solution of nominal concentration $x^0_{\textrm{Sc}}=0.005$
at $T=773$~K: evolution with time of the number $N_{sp}$ of stable
precipitates in the simulation box (normalized by the number of
lattice sites $N_s$), of stable precipitate average size
$<n_{\textrm{Sc}}>_{sp}$, and of Sc concentration
$x_{\textrm{Sc}}$ in the solid solution. The critical size used to
discriminate stable precipitates from sub-critical clusters is
$n^*_{\textrm{Sc}}=13$. Some of the corresponding simulation
configurations are shown in Fig.~\ref{alsc_kinetic_picture}.}
\label{alsc_kinetic_fig}
\end{figure}

So as to follow kinetics of precipitation in the simulation box,
we have to give us a criterion to discriminate atoms belonging to
the solid solution from those in L1$_2$ precipitates. As,
according to the phase diagram, the stoichiometry of theses
precipitates is almost perfect, we only look at Zr or Sc atoms and
assume for each of these atoms in a L1$_2$ cluster that three
associated Al atoms belong to the same cluster. Zr (or Sc) atoms
are counted as belonging to a cluster having L1$_2$ structure if
all their twelve first nearest neighbors are Al atoms and at least
one of their six second nearest neighbors is a Zr (or Sc) atom.
This criterion works only for dimers and bigger clusters and then
all Zr or Sc atoms not belonging to such clusters are considered
to be monomers. We only counted as precipitates L1$_2$ clusters
bigger than a critical size, \ie containing more Zr or Sc atoms
than a critical number $n^*_{\textrm{X}}$, this critical number
being chosen as the initial one given by classical nucleation
theory (\cf section~\ref{cnt_chap}). Clusters smaller than this
critical size are unstable and will re-dissolve into the solid
solution. Therefore atoms contained in such clusters are counted
as belonging to the solid solution. With this criterion, we can
measure during the atomic simulations the number of stable
precipitates and their mean size (Fig.~\ref{alsc_kinetic_picture}
and \ref{alsc_kinetic_fig}). The solid solution concentration is
then defined at each step by the relation
\begin{equation}
x_{\textrm{X}}=\sum_{\nx=1}^{\nx^*}{\nx C_{\nx}}, \label{conc_def}
\end{equation}
where $C_{\nx}$ is the number of L1$_2$ clusters containing $\nx$
Zr or Sc atoms normalized by the number of lattice sites, \ie the
instantaneous probability to observe such a cluster in the
simulation box.

All starting configurations for simulations are completely
disordered (random) solid solutions. We thus simulate infinitely
fast quenching from high temperatures. During the first steps of
the precipitation, the number of stable precipitates is varying
quite linearly with time (Fig.~\ref{alsc_kinetic_fig}). The slope
of this linear relation gives a measure of the steady-state
nucleation rate $J^{st}$, \ie the number of stable precipitates
appearing by time unit
during the nucleation stage.
\section{Classical nucleation theory}
\label{cnt_chap}

In order to compare kinetic Monte Carlo simulations with classical
nucleation theory, we need to define the formation free energy
$\Delta G_n(x^0_{\textrm{X}})$ of a L1$_2$ cluster containing $n$
atoms ($n = 4n_{\textrm{Zr}}$ or $4n_{\textrm{Sc}}$) embedded in a
solid solution of nominal concentration $x^0_{\textrm{X}}$.
Usually, one uses the capillary approximation and considers a
volume contribution, the nucleation free energy $\Delta
G^{nuc}(x^0_{\textrm{X}})$, and a surface contribution
corresponding to the energy cost to create an interface between
the solid solution and the L1$_2$ precipitate,
\begin{equation}
\Delta G_n(x^0_{\textrm{X}}) = n \Delta G^{nuc}(x^0_{\textrm{X}})
+ \left(\frac{9\pi}{4}\right)^{1/3} n^{2/3} a^2 \bar{\sigma}
\label{Gn_eq} ,
\end{equation}
where $a$ is the lattice parameter and $\bar{\sigma}$ the
interface free energy.  For a supersaturated solution, $\Delta
G_n(x^0_{\textrm{X}})$ shows a maximum in $n^* =
4n^*_{\textrm{Zr}}$ or $4n^*_{\textrm{Sc}}$ corresponding to the
critical size used to follow the kinetics of precipitation during
the Monte Carlo simulations (\cf section~\ref{KMC_chap}). We now
have to calculate the nucleation free energy $\Delta
G^{nuc}(x^0_{\textrm{X}})$ and the interface free energy
$\bar{\sigma}$ corresponding to the set of atomic parameters
presented in the previous section.

\subsection{Nucleation free energy}
\label{Gnuc_chap}

The nucleation free energy to precipitate Al$_3$X (X~$\equiv$ Zr
or Sc) is \cite{MAR78,PORTER}
\begin{eqnarray}
\Delta G^{nuc}(x^0_{\textrm{X}}) &=&
  \frac{3}{4} \left( \mu_{\textrm{Al}}(x^{eq}_{\textrm{X}})
 - \mu_{\textrm{Al}}(x^{0}_{\textrm{X}}) \right)
\nonumber\\
&&+ \frac{1}{4} \left( \mu_{\textrm{X}}(x^{eq}_{\textrm{X}}) -
\mu_{\textrm{X}}(x^{0}_{\textrm{X}}) \right) , \label{Gnuc}
\end{eqnarray}
where $\mu_{\textrm{Al}}(x_{\textrm{X}})$ and
$\mu_{\textrm{X}}(x_{\textrm{X}})$ are the chemical potentials of
respectively Al and X components in the solid solution of
concentration $x_{\textrm{X}}$, $x^{eq}_{\textrm{X}}$ is the
equilibrium concentration of the solid solution, and
$x^{0}_{\textrm{X}}$ the nominal concentration. The factors $3/4$
and $1/4$ arise from the stoichiometry of the precipitating phase
Al$_3$X. We use the CVM in the tetrahedron-octahedron
approximation \cite{KIK51,SAN78} to calculate chemical potentials
entering Eq.~(\ref{Gnuc}). This is the minimum CVM approximation
that can be used with first and second nearest neighbor
interactions. Within this approximation, all correlations inside
the tetrahedron of first nearest neighbors and the octahedron
linking the centers of the six cubic faces are included in the
calculation of the chemical potentials. Usually one does not
consider these correlations in the calculation of the nucleation
free energy and merely uses the Bragg-Williams approximation to
obtain $\Delta G^{nuc}$ but we will see in section
\ref{mean_field_chap} that this leads to discrepancies between
results of atomic simulations and predictions of classical
nucleation theory.

\subsection{Interface free energy}
\label{sigma_chap}

\subsubsection{Plane interfaces}

\begin{figure}
\includegraphics[width=0.9\linewidth]{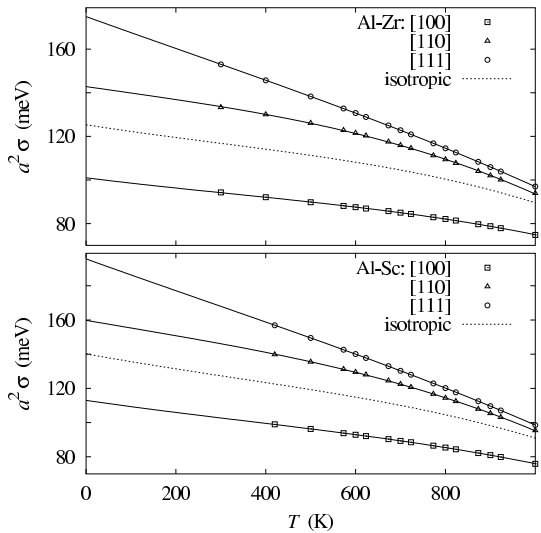}
\caption{Dependence with temperature of the free energies of the
solid solution~/ Al$_3$Zr (top) and solid solution~/ Al$_3$Sc
(bottom) interfaces for the [100], [110], and [111] directions,
and associated isotropic free energy $\bar{\sigma}$ obtained from
Wulff construction.} \label{interface_fig}
\end{figure}

We calculate interface free energies between the aluminum solid
solution assumed to be at equilibrium close to the interface and
the L1$_2$ precipitates for three different directions of the
interface ([100], [110], and [111]). If phases are assumed to be
pure, the different interface energies are simply related by the
equation
\begin{equation}
\sigma_{100} = \frac{1}{\sqrt{2}} \sigma_{110} =
\frac{1}{\sqrt{3}} \sigma_{111} = \frac{\omega^{(2)}}{a^2} .
\label{sigma0K}
\end{equation}
At finite temperature, one has to consider that the solid solution
is not pure Al and that the L1$_2$ structure differs from Al$_3$X
stoichiometry. Moreover, to minimize the energy cost due to the
interface, concentrations and order parameters of planes near the
interface can differ from those in the bulk. So as to take into
account such a relaxation, we calculate these interface free
energies within the Bragg-Williams approximation. A better
statistical approximation based on CVM is too cumbersome and we
only check for the [100] direction that we obtain the same value
of the free energy in the whole range of temperatures with a CVM
calculation in the tetrahedron approximation.

\begin{figure}
\subfigure[$T=723$~K]{
\includegraphics[width=0.60\linewidth]{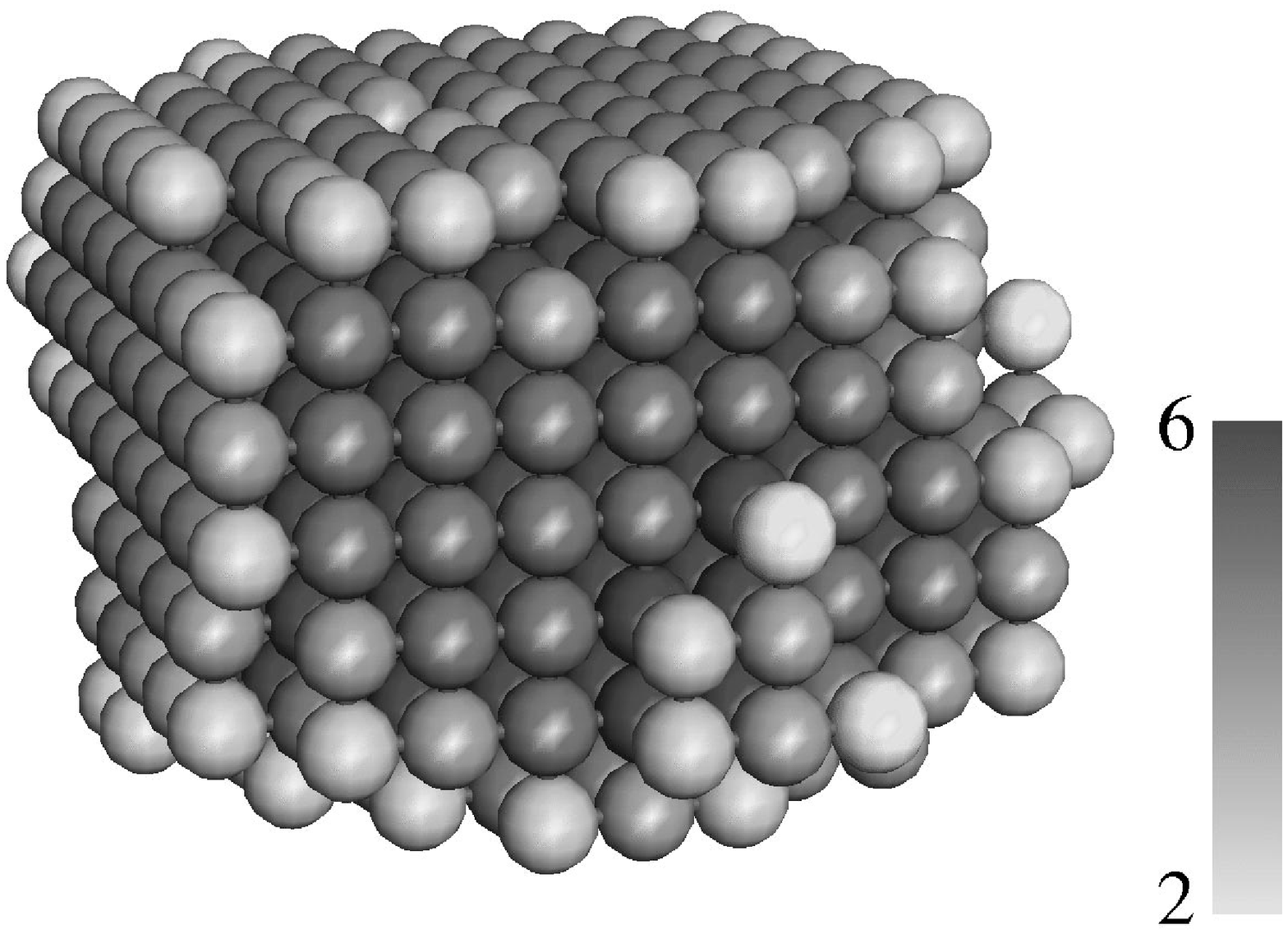}
\hfill
\includegraphics[width=0.20\linewidth]{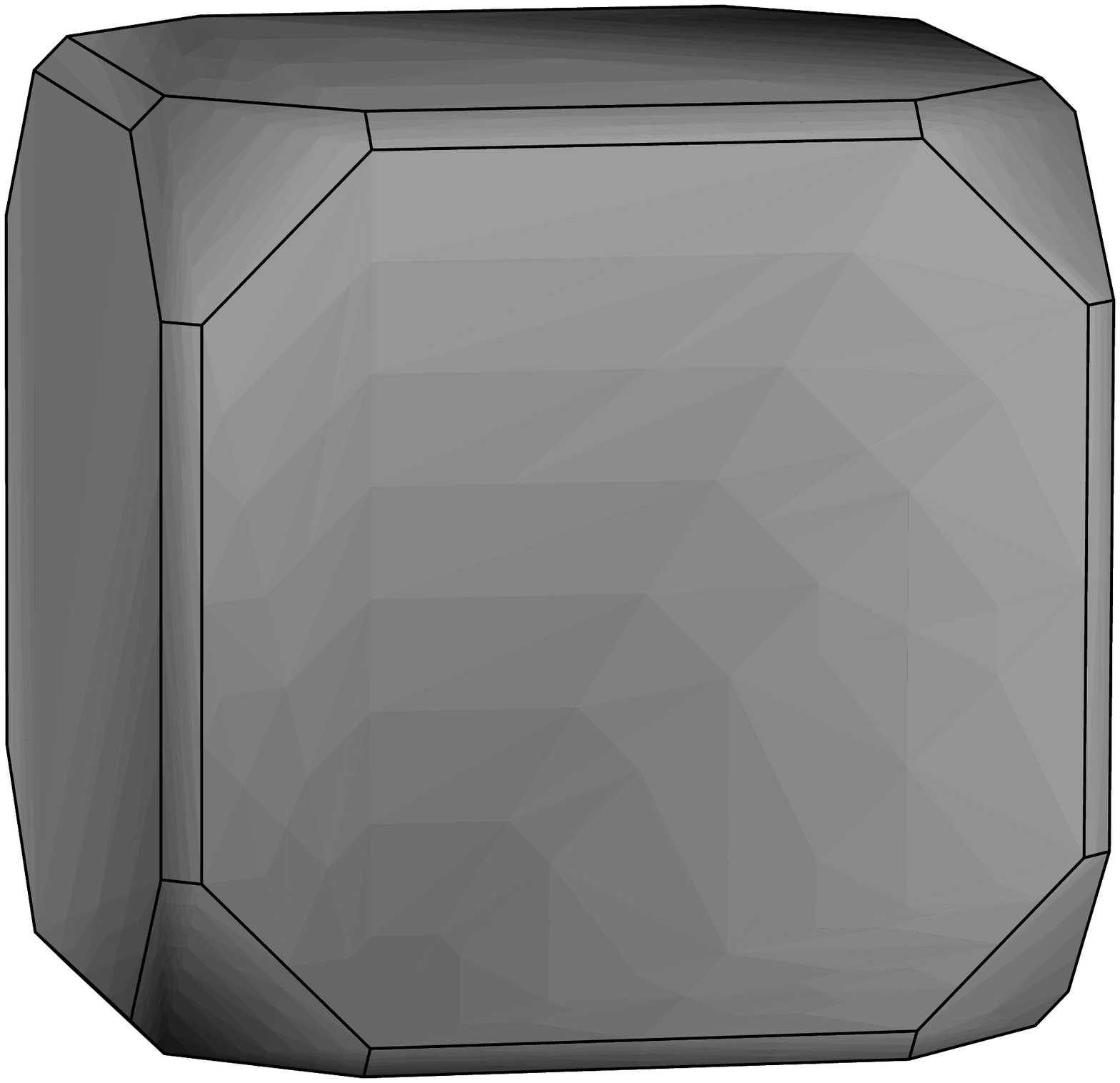}}
\subfigure[$T=873$~K]{
\includegraphics[width=0.60\linewidth]{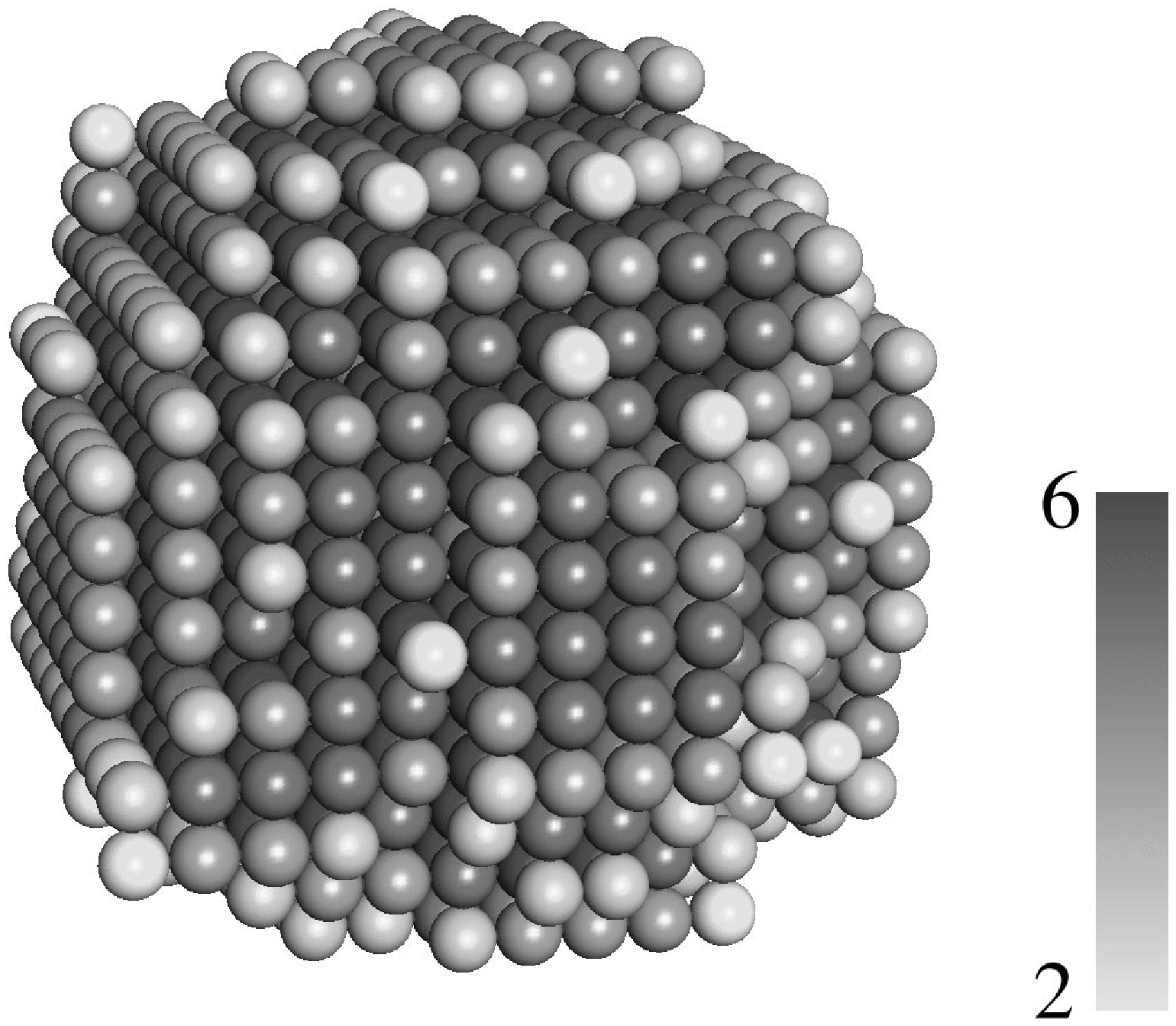}
\hfill
\includegraphics[width=0.20\linewidth]{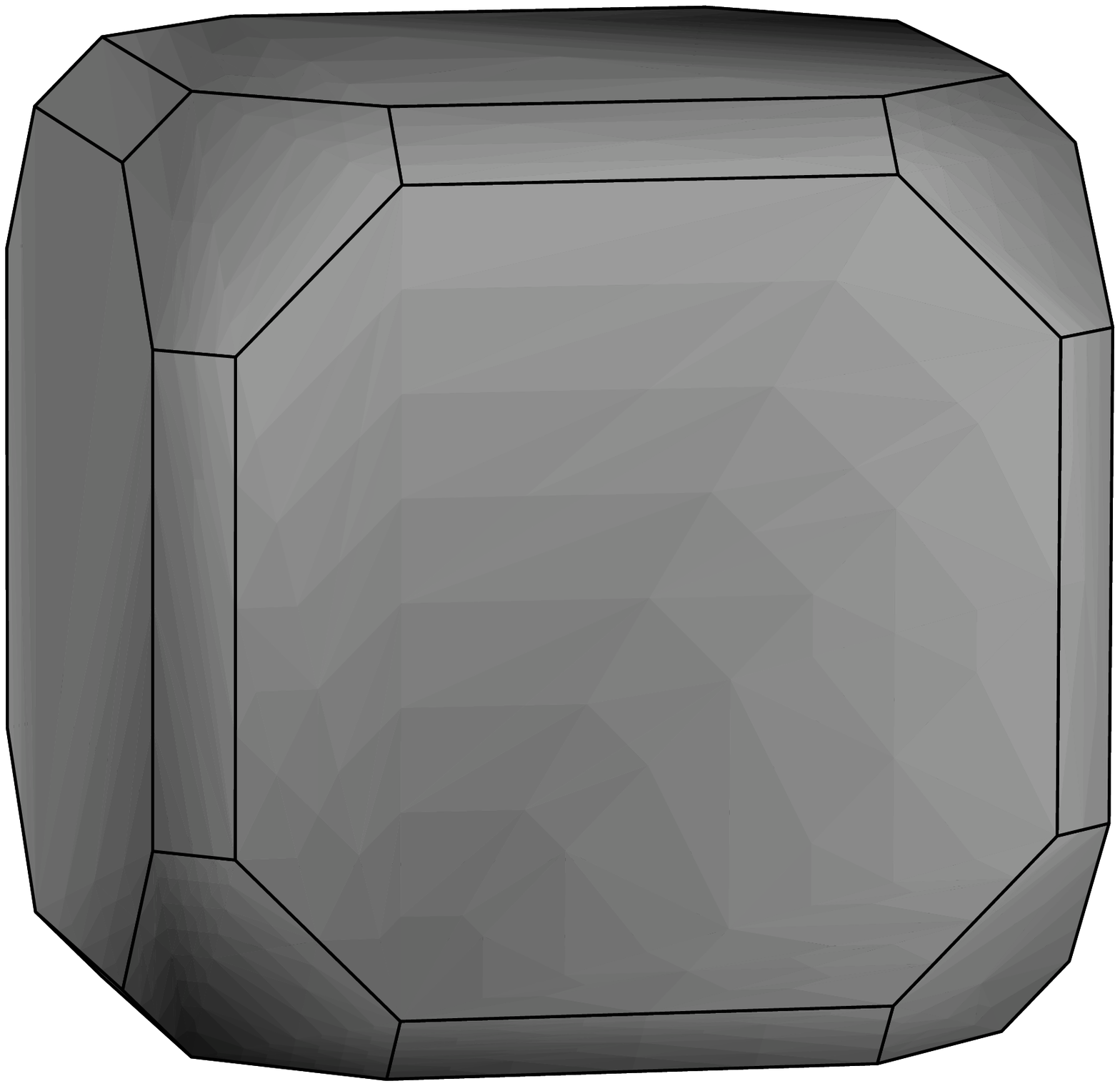}}
\caption{Al$_3$Sc precipitate observed during Monte Carlo
simulations at different temperatures and corresponding Wulff
construction obtained from the interface free energies calculated
at the same temperatures. For Monte Carlo simulations, only Sc
atoms are shown. Atom color corresponds to the number of Sc atoms
as second nearest neighbors. For a [100] interface, it should be
5, for a [110] 4, and for a [111] 3.} \label{interface_picture}
\end{figure}

At finite temperature, we still observe that $\sigma_{100} <
\sigma_{110} < \sigma_{111}$ (Fig.\ref{interface_fig}).
Nevertheless, as the relaxation is small for the interface in the
[100] direction and important for the [111] direction, the
difference between interface energies is decreasing with
temperature. This indicates that precipitates are becoming more
isotropic at higher temperatures. Using Wulff construction
\cite{PORTER,CHRISTIAN} to determine the precipitate equilibrium
shape, we find that precipitates will mainly show facets in the
[100] directions and that facets in the [110] and [111] directions
are small but becoming more important with increasing temperature.
Comparing these predicted equilibrium shape with the ones observed
during the atomic simulations, we find a good agreement
(Fig.~\ref{interface_picture}): at low temperatures
($T\sim723$~K), precipitates are cubic with sharp [100]
interfaces, whereas at higher temperatures interfaces are not so
sharp. For Al$_3$Sc, Marquis and Seidman \cite{MAR01}
experimentally observed precipitates showing facets in the [100],
[110], and [111] directions at $T=573$~K, with [100] facets
tending to disappear at high temperatures. This is well reproduced
by our atomic model, the main difference being that the
experimentally observed [100] facets are less important compared
to the other ones than in our study.

Asta \etal \cite{AST98} used a cluster expansion of ab-initio
calculations to obtain the same interface energies in Al-Sc
system. The energies they got are higher than ours:
$a^2\sigma_{100}$ is varying from 167 to 157~meV between 0K and
the melting temperature ($T^{mel}=934$K) and $a^2\sigma_{111}$
from 233 to 178~meV. The difference could be due to the limited
range of our interactions compared to Asta's ones. This could
explain too why the interface free energies we obtain are
decreasing more rapidly with temperature especially in the [111]
direction \cite{SLU96}.

Hyland \etal \cite{HYL98} also calculated interface free energies
with an empirical potential for Al-Sc, but the energies they
obtained are really low compared to ours and Asta's ones as well
as compared to the isotropic interface free energy they measured
\cite{HYL92}. Some of the discrepancy can be due to relaxations of
atomic positions which are considered in their study and are
missing in our. It may also indicate that the potential they used
is not really well suited to describe solid solution~/ Al$_3$Sc
interfaces.

\subsubsection{Average interface free energy}

We use the Wulff construction \cite{PORTER,CHRISTIAN}
 to define an isotropic free energy $\bar{\sigma}$
from the free energies $\sigma_{100}$, $\sigma_{110}$, and
$\sigma_{100}$. $\bar{\sigma}$ is defined so as to give the same
interface free energy for a spherical precipitate having the same
volume as the real faceted one. Details of calculations can be
found in appendix \ref{Wulff_chap}. The free energy $\bar{\sigma}$
is higher than the minimum energy $\sigma_{100}$
(Fig.~\ref{interface_fig}). The ratio $\bar{\sigma}/\sigma_{100}$
is slightly lower than $(6/\pi)^{1/3}$, this value corresponding
to cubic precipitates showing only [100] facets.

Robson and Prangnell \cite{ROB01} deduced from experimental
observations of Al$_3$Zr coarsening a Al/Al$_3$Zr interface free
energy $\bar{\sigma}=100$~mJ.m$^{-2}$ at 773~K. The agreement
between this value and the one deduced from our atomic model is
perfect.
In the same way, Hyland \cite{HYL92} obtained from measured
nucleation rates and incubation times an experimental Al/Al$_3$Sc
interface energy $\bar{\sigma}=94\pm23$~mJ.m$^{-2}$ between 563
and 623~K. Nevertheless, this experimental value should be
considered only as an order of magnitude as experimental
nucleation rates and incubation times are hard to obtain. One has
to be sure that precipitates of the critical size can be observed
and the difference between the interface energies deduced from the
incubation times or from the nucleation rates could be due to a
detection limit for small precipitates greater than the critical
size. Moreover, the Sc diffusion coefficient used by Hyland in his
study differs from the one which has been more recently obtained
from radioactive tracer diffusion measurements \cite{FUJ97} and
this would influence too the value of the interface energy deduced
from his experimental observations. With these considerations in
mind, this experimental value although slightly lower than the one
we calculate ($\sim$113~mJ.m$^{-2}$) is in good agreement with it.
This indicates that the use of Wulff construction with mean-field
theory is a good way to estimate this isotropic interface free
energy and that our set of atomic parameters
(Tab.~\ref{energies_tab}) is realistic to model solid solution~/
Al$_3$Zr and Al$_3$Sc interfaces.

\subsection{Cluster size distribution}

For a dilute solution, the probability to observe in the solid
solution a cluster containing $n$ atoms having L1$_2$ structure is
\cite{FRENKEL,MAR78}
\begin{equation}
C_n \sim \frac{C_n}{1-\sum_j C_j} = \exp{\left( - \Delta G_n / kT
\right)}, \label{distri_cnt}
\end{equation}
where the formation energy $\Delta G_n$ is given by
Eq.~(\ref{Gn_eq}). If the solution is supersaturated, the energy
$\Delta G_n$ is decreasing for sizes greater than the critical
size and Eq.~(\ref{distri_cnt}) is assumed to be checked only for
$n\leq n^*$. As this is the criterion we chose to discriminate the
solid solution from the L1$_2$ precipitates (\cf section
\ref{KMC_chap}), this means that only the cluster size
distribution in the solid solution should obey
Eq.~(\ref{distri_cnt}) and not the size distribution of stable
precipitates.

\begin{figure}
\includegraphics[width=0.99\linewidth]{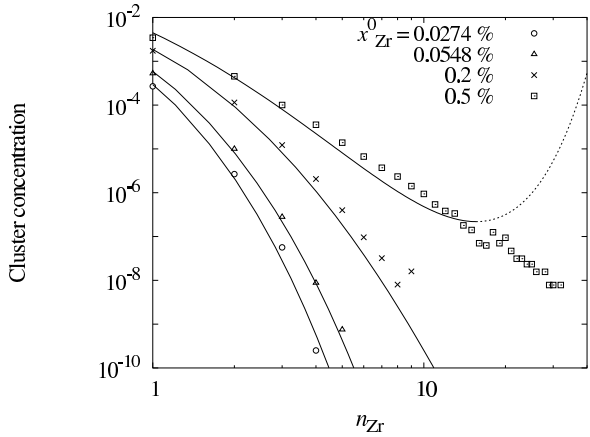}
\caption{Dependence with the nominal concentrations
$x^0_{\textrm{Zr}}$ of the cluster size distributions of an
aluminum solid solution at $T=773$~K. At this temperature, the
solubility limit is $x^{eq}_{\textrm{Zr}}=5.48\times10^{-4}$.
Lines correspond to prediction of classical nucleation theory
combined with CVM calculation and symbols to Monte Carlo
simulations.} \label{distri1_fig}
\end{figure}

We compare the cluster size distribution given by the
Eq.~(\ref{distri_cnt}) with the ones measured in Monte Carlo
simulations for different temperatures between 723 and 873~K and
different concentrations of the solid solution in the Al-Zr
(Fig.~\ref{distri1_fig}) as well as Al-Sc systems, both systems
leading to the same conclusions.

For stable solid solutions ($x^0_{\textrm{X}} <
x^{eq}_{\textrm{X}}$), all energetic contributions entering
$\Delta G_n$ are positive and the cluster critical size is not
defined. Therefore, one expects Eq.~(\ref{distri_cnt}) to be
obeyed for all values of $n$. A comparison with Monte Carlo
simulations shows a good agreement. The comparison can only be
made for small clusters: the probability to observe large clusters
in the simulations is too low to obtain statistical information on
their distribution in a reasonable amount of computational time.
This is interesting to note that for a solid solution having a
concentration equal to the solubility limit ($x^0_{\textrm{X}} =
x^{eq}_{\textrm{X}}$), the prediction \ref{distri_cnt} of the
cluster size distribution is still correct. As the nucleation free
energy is null for this concentration, the only contribution to
$\Delta G_n$ arises from the interface. This shows that our
estimation of the interface free energy $\bar{\sigma}$ is coherent
with its use in Eq.~(\ref{distri_cnt}) and that the capillary
approximation gives a good description of the solid solution
thermodynamics.

For low supersaturated solid solution (for instance, on
Fig.~\ref{distri1_fig}, $x^0_{\textrm{Zr}}=0.2$~\%), we observe a
stationary state during Kinetic Monte Carlo simulations: the
computational time to obtain a stable L1$_2$ cluster is too high
and the solid solution remains in its metastable state. Therefore,
we can still measure the cluster size distribution during the
simulations. The agreement with Eq.~(\ref{distri_cnt}) is still
correct (Fig.~\ref{distri1_fig}). One should notice that now, the
critical size being defined, the comparison is allowed only for
$n\leq n^*$.

\begin{figure}
\includegraphics[width=0.99\linewidth]{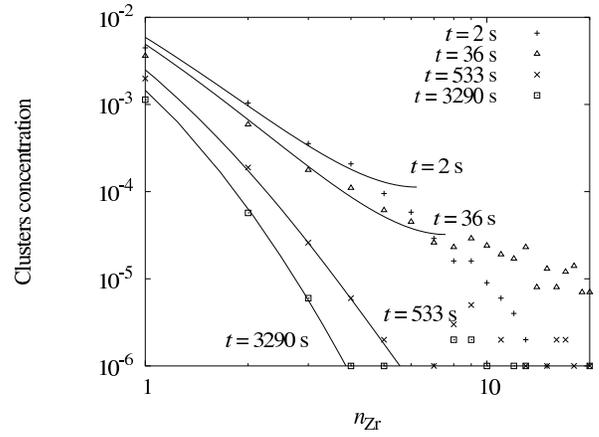}
\caption{Evolution with time of the cluster size distributions of
an aluminum solid solution of nominal concentration
$x^0_{\textrm{Zr}}=1\times10^{-2}$ at $T=723$~K. At this
temperature, the solubility limit is
$x^{eq}_{\textrm{Zr}}=2.90\times10^{-4}$. Symbols correspond to
Monte Carlo simulations and lines to prediction of classical
nucleation theory combined with CVM calculation with the following
instantaneous solid solution concentrations and critical sizes:
$x_{\textrm{Zr}}=1\times10^{-2}$, $7\times10^{-3}$,
$2.7\times10^{-3}$, and $1.5\times10^{-3}$ and
$n^*_{\textrm{Zr}}=7$, 8, 18, and 41 at respectively $t=2$, 36,
533, and 3290~s.} \label{distri2_fig}
\end{figure}

For higher supersaturations, the solid solution concentration
$x_{\textrm{X}}$ is decreasing meanwhile stable precipitates
appear (\cf kinetics of precipitation of Al$_3$Sc in
Fig.~\ref{alsc_kinetic_fig}). This involves that the nucleation
free energy is decreasing in absolute value and that the critical
size $n^*$ is increasing. At each step we have to re-calculate the
solid solution concentration and the critical size
self-consistently by means of the definition (\ref{conc_def}) of
$x_{\textrm{X}}$ and by imposing that $\Delta G_n(x_{\textrm{X}})$
is maximum in $n^*$. Then we use this new value of the solid
solution concentration in Eq.~(\ref{distri_cnt})  to calculate the
corresponding cluster size distribution and compare it with the
kinetic Monte Carlo simulation (\cf cluster size distributions in
a Zr supersaturated aluminum solution in Fig.~\ref{distri2_fig}).
We see that the time evolution of the cluster size distribution is
well reproduced by Eq.~(\ref{distri_cnt}) when the instantaneous
concentration is used to calculate the nucleation free energy and
therefore the prediction of cluster size distribution is not only
verified during the nucleation stage but is well adapted even
during the growing stage. Thus the thermodynamic description used
in the classical nucleation theory is in good agreement with
results of atomic simulations.

\subsection{Kinetic description}
\subsubsection{Diffusion}
\label{diffusion_chap}

Classical nucleation theory assumes that only monomers migrate and
that larger clusters like dimers do not diffuse. We check this is
the case with our atomic model by measuring during Monte Carlo
simulations the diffusion coefficient associated to the gravity
center of $N$ atoms X in pure Al for $2 \leq N \leq 4$. We obtain
for Zr as well as Sc atoms that this diffusion coefficient is
equal to the monomer diffusion coefficient divided by the number
$N$ of X atoms considered. This implies the following relation
\begin{equation}
\left< \left( \sum_{n=1}^{N}{ \mathbf{\Delta r}_{\textrm{X}_n} }
\right)^2 \right> = \sum_{n=1}^{N}{ \left< {\mathbf{\Delta
r}_{\textrm{X}_n}}^2 \right> } , \label{diff_correlation_eq}
\end{equation}
where the brackets indicate a thermodynamic ensemble average and
$\mathbf{\Delta r}_{\textrm{X}_n}$ is the displacement of the atom
X$_n$ during a given time. This relation is satisfied only if
there is no correlation between the displacement of the $N$ atoms
X, which in other words means that cluster formed of the $N$ atoms
does not diffuse. In both systems, tracer diffusion coefficient of
Al is several order of magnitude larger than tracer diffusion
coefficients of X. The relationship of Manning \cite{MAN71} shows
that in that case the correlation factor $f_{\textrm{XX}}$ is
almost equal to the tracer correlation factor $f_{\textrm{X}}$
which is equivalent\cite{ALL82,ALLNATT} to
Eq.~(\ref{diff_correlation_eq}). Thus the assumption used by
classical nucleation theory of a diffusion controlled by monomers
is checked for both Al-Zr and Al-Sc systems although interactions
with vacancies are different for these two binary systems. This is
not the case for all systems: when vacancies are trapped inside
precipitates or at the interface with the matrix, small clusters
can migrate \cite{SOI00,ATH00,ROU01,LEB02} which affects kinetics
of precipitation.

\subsubsection{Steady-state nucleation rate}

\begin{figure}
\includegraphics[width=0.99\linewidth]{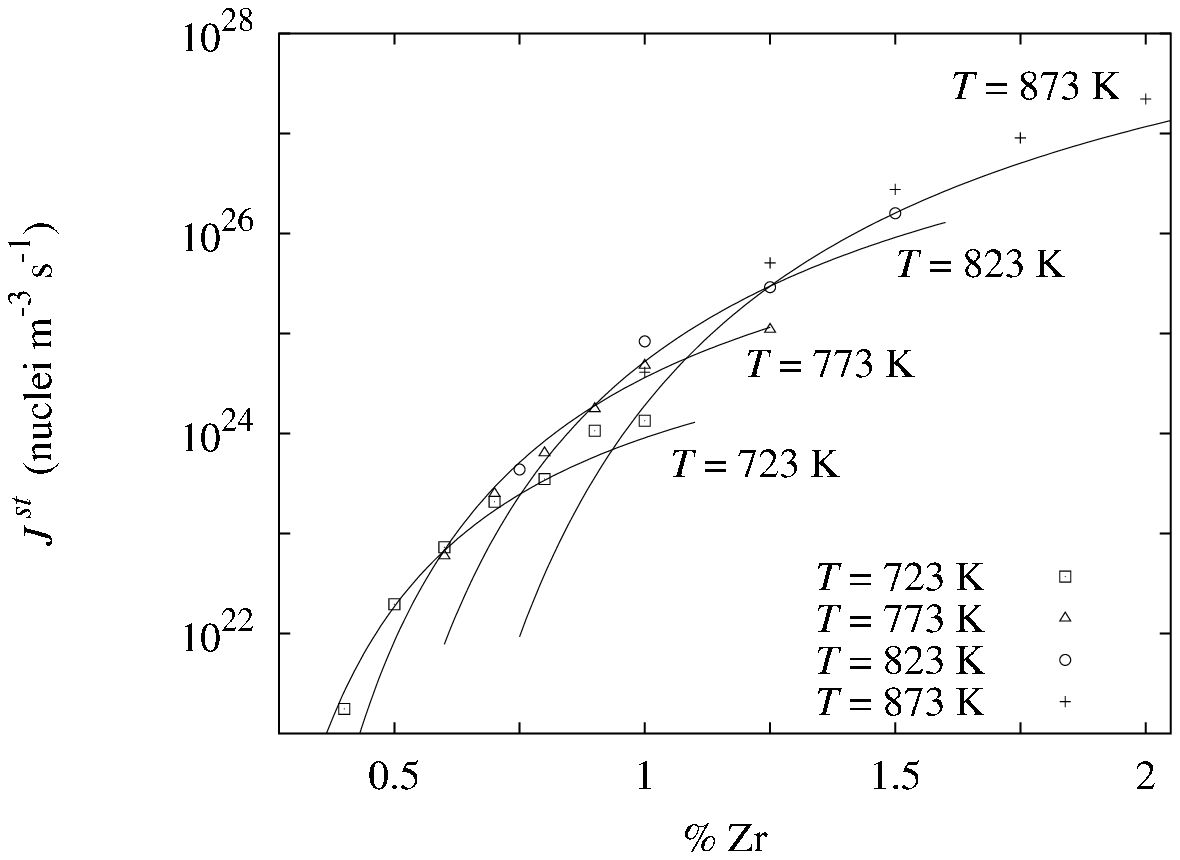}
\includegraphics[width=0.99\linewidth]{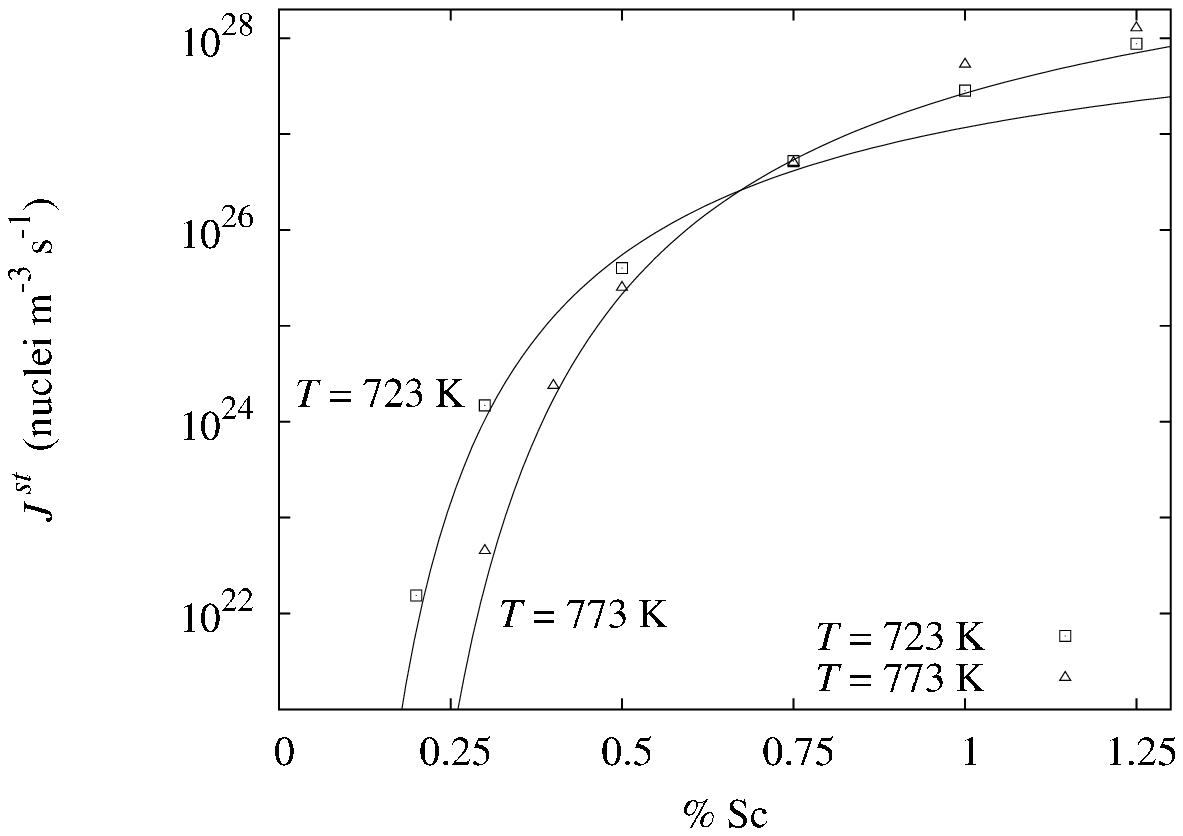}
\caption{Variation with nominal concentration and temperature of
the steady-state nucleation rate $J^{st}$ for Al$_3$Zr (top) and
Al$_3$Sc (bottom) precipitation. Symbols correspond to Monte Carlo
simulations and lines to classical nucleation theory combined with
CVM calculation.} \label{Jst_fig}
\end{figure}

The steady-state nucleation rate is then predicted to be given by
the equation \cite{MAR78},
\begin{equation}
J^{st} = N_s Z \beta^* \exp{ \left(- \Delta G^* /kT \right) },
\label{nucleation_rate_eq}
\end{equation}
where $N_s$ is the number of nucleation sites, \ie the number of
lattice sites, $\Delta G^*$ is the nucleation barrier and
corresponds to the free energy of a precipitate of critical size
$n^*$ as given by Eq.~(\ref{Gn_eq}),
$Z$ is the Zeldovitch factor and describes size fluctuations of
precipitates around $n^*$,
\begin{equation}
Z= \frac{\left(\Delta {G^{nuc}}\right)^2}{2\pi (a^2
\bar{\sigma})^{3/2} \sqrt{kT}} \label{Zeldo_eq},
\end{equation}
and $\beta^*$ is the condensation rate for clusters of critical
size $n^*$. Assuming the limiting step of the adsorption is the
long range diffusion of Zr or Sc in the solid solution and that Al
atoms diffuse infinitely faster than Zr or Sc atom, the
condensation rate is \cite{MAR78}
\begin{equation}
\beta^* = - 32 \pi \frac{a^2 \bar{\sigma}}{\Delta G^{nuc}}
\frac{D_{\textrm{X}}}{a^2} x^0_{\textrm{X}}.
\label{condensation_rate}
\end{equation}
Although only monomers diffuse, the concentration appearing in
Eq.~(\ref{condensation_rate}) is the nominal one as it reflects
the gradient of concentration driving diffusion. Each time one Zr
or Sc atom condensates on a cluster, three Al atoms condensate too
on the same cluster. Thus clusters are growing from sizes $4n$ to
$4(n+1)$.

Comparing with the steady-state nucleation rate measured in Monte
Carlo simulations for different temperatures and different
supersaturations of the solid solution in the Al-Zr and Al-Sc
system, we see that the classical nucleation theory manages to
predict $J^{st}$ (Fig.~\ref{Jst_fig}). The agreement is really
good for low nominal concentrations of the solid solution
($x_{\textrm{X}} \leq 1\times10^{-2}$) and is still good for
higher concentrations: there is a small discrepancy but the
relative values for different temperatures at a given
concentration are correctly predicted. For instance, for the
nominal concentration $x^0_{\textrm{Zr}}=0.01$, we obtain that the
steady-state nucleation rate is higher at $T=823$~K than at
$T=773$ or 873~K. This shows that the kinetic model used by the
classical nucleation theory is checked both for Al$_3$Zr kinetics
of precipitation where there is repulsion between the vacancy and
the precipitating element and for Al$_3$Sc kinetics
where there is attraction.

\section{Capillary approximation}

Although it manages to catch thermodynamics of the solid solution,
the capillary approximation that we used previously can look
rough. First of all, one can wonder if it is reasonable to assume
spherical precipitates especially for small ones. Moreover, when
counting precipitates in Monte Carlo simulations, we assume them
as being stoichiometric whereas in the mean-field calculation of
the nucleation free energy we include anti-site defect
contribution. Another source of mistake could be the use of the
Wulff construction to calculate an isotropic interface free
energy: doing so, we calculate the interface free energy of the
most stable precipitate and therefore neglect some configurational
entropy.

In the present section, we calculate the cluster free energy
without using the capillary approximation. The results obtained
with this \emph{direct} calculation are then confronted to the
ones obtained with the capillary approximation. We also take the
benefit of the exact results to discuss different levels of
mean-field approximation used for the calculation of parameters
entering in the capillary approximation.

\subsection{Direct calculation of cluster energies.}

Instead of using the capillary approximation to calculate the
formation energy of L1$_2$ clusters, we can calculate this
quantity exactly. This can be done, following
Ref.~\onlinecite{PER84}, by sampling thermodynamic averages with
Monte Carlo simulations so as to compute the free energy
difference between a cluster of size $n$ and one of size $n+1$ at
a given temperature. This method presents the drawback that a
calculation is needed at every temperature of interest. We prefer
calculating all coefficients entering the partition function as
done in Ref.~\onlinecite{BER03} and then derive the free energy at
every temperature.

A L1$_2$ cluster containing $n_{\textrm{X}}$ X atoms can have
different shapes which we group by classes $\alpha$ of same
energy: $D_{n_{\textrm{X}},\alpha}$ is the number per lattice site
of clusters containing $n_{\textrm{X}}$ X atoms and having the
energy 
$H_{n_{\textrm{X}},\alpha} = n_{\textrm{X}} \left(
12\omega^{(1)} + 6 \epsilon_{\textrm{XX}}^{(1)} - 6
\epsilon_{\textrm{AlAl}}^{(1)} + 3 \epsilon_{\textrm{XX}}^{(2)} -
3 \epsilon_{\textrm{AlAl}}^{(2)} \right) + \delta
H_{n_{\textrm{X}},\alpha}$.
Energies are defined
referred to the pure Al reference state as the cluster energy
is the energy change due to the presence of a cluster in pure Al.
The free energy of a L1$_2$ cluster containing $n_{\textrm{X}}$ X
atoms is then defined by
\begin{eqnarray}
G_{n_{\textrm{X}}} &=& -kT \ln{\left( \sum_{\alpha}
D_{n_{\textrm{X}},\alpha}
\exp{ \left( - H_{n_{\textrm{X}},\alpha} /kT \right) } \right)} \nonumber\\
&=& n_{\textrm{X}} \left( 12\omega^{(1)} + 6
\epsilon_{\textrm{XX}}^{(1)} - 6 \epsilon_{\textrm{AlAl}}^{(1)} +
3 \epsilon_{\textrm{XX}}^{(2)} - 3 \epsilon_{\textrm{AlAl}}^{(2)}
\right) \nonumber \\
&& -kT \ln{\left( \sum_{\alpha} D_{n_{\textrm{X}},\alpha} \exp{
\left( - \delta H_{n_{\textrm{X}},\alpha} /kT \right) } \right)}
\label{direct_partition_eq}
\end{eqnarray}
Degeneracies $D_{n_{\textrm{X}},\alpha}$ can be computed for a
given size by generating clusters with a random configuration and
then by counting for each energy level $\alpha$ the number of
different clusters. The obtained values are presented in table
\ref{direct_tab} for L1$_2$ clusters containing less than 9 X
atoms. For bigger clusters, the degeneracy of the different
classes is becoming too high to be countable. This is important to
notice that we use the same criterion to define L1$_2$ clusters as
in the kinetic Monte Carlo simulations (\cf
section~\ref{KMC_chap}), and that Zr or Sc atoms belonging to a
L1$_2$ cluster only have Al atoms as first nearest neighbors. We
thus do not allow anti-site defects on the majority sub-lattice.
This is not an important restriction as these defects have a high
formation energy and therefore their contribution to the partition
function can be neglected. As for the minority sub-lattice,
anti-site defects can not be taken into account as they lead to a
change of the precipitate size.

\begin{table}[!hbtp]
\caption{Degeneracies $D_{n_{\textrm{X}},\alpha}$ corresponding to
classes of L1$_2$ clusters containing $n_{\textrm{X}}$ X atoms and
having energy $H_{n_{\textrm{X}},\alpha} = n_{\textrm{X}} \left(
12\omega^{(1)} + 6 \epsilon_{\textrm{XX}}^{(1)} - 6
\epsilon_{\textrm{AlAl}}^{(1)} + 3 \epsilon_{\textrm{XX}}^{(2)} -
3 \epsilon_{\textrm{AlAl}}^{(2)} \right) + \delta
H_{n_{\textrm{X}},\alpha}$ for $1 \leq n_{\textrm{X}} \leq 9$.}
\label{direct_tab}
\begin{ruledtabular}
\begin{tabular}{rrcrcrrcr}
\multicolumn{1}{c}{$n_{\textrm{X}}$} &
\multicolumn{1}{c}{$\alpha$} & \multicolumn{1}{c}{$\delta
H_{n_{\textrm{X}},\alpha}$}&
\multicolumn{1}{c}{$D_{n_{\textrm{X}},\alpha}$} &&
\multicolumn{1}{c}{$n_{\textrm{X}}$} &
\multicolumn{1}{c}{$\alpha$} & \multicolumn{1}{c}{$\delta
H_{n_{\textrm{X}},\alpha}$}&
 \multicolumn{1}{c}{$D_{n_{\textrm{X}},\alpha}$} \\
\hline
&&&                                && 7 & 1 & $ 24 \omega^{(2)} $ & 8 \\
1 & 1 & $  6 \omega^{(2)} $ & 1    && 7 & 2 & $ 26 \omega^{(2)} $ & 378 \\
&&&                                && 7 & 3 & $ 28 \omega^{(2)} $ & 4368 \\
2 & 1 & $ 10 \omega^{(2)} $ & 3    && 7 & 4 & $ 30 \omega^{(2)} $ & 18746 \\
\\
3 & 1 & $ 14 \omega^{(2)} $ & 15   && 8 & 1 & $ 24 \omega^{(2)} $ & 1 \\
&&&                                && 8 & 2 & $ 28 \omega^{(2)} $ & 306 \\
4 & 1 & $ 16 \omega^{(2)} $ & 3    && 8 & 3 & $ 30 \omega^{(2)} $ & 4829 \\
4 & 2 & $ 18 \omega^{(2)} $ & 83   && 8 & 4 & $ 32 \omega^{(2)} $ & 35926 \\
&&&                                && 8 & 5 & $ 34 \omega^{(2)} $ & 121550 \\
5 & 1 & $ 20 \omega^{(2)} $ & 48   && \\
5 & 2 & $ 22 \omega^{(2)} $ & 486  && 9 & 1 & $ 28 \omega^{(2)} $ & 24 \\
&&&                                && 9 & 2 & $ 30 \omega^{(2)} $ & 159 \\
6 & 1 & $ 22 \omega^{(2)} $ & 18   && 9 & 3 & $ 32 \omega^{(2)} $ & 5544 \\
6 & 2 & $ 24 \omega^{(2)} $ & 496  && 9 & 4 & $ 34 \omega^{(2)} $ & 51030 \\
6 & 3 & $ 26 \omega^{(2)} $ & 2967 && 9 & 5 & $ 36 \omega^{(2)} $ & 289000 \\
&&&                                && 9 & 6 & $ 38 \omega^{(2)} $ & 803000 \\
\end{tabular}
\end{ruledtabular}
\end{table}

The formation energies entering Eq.~(\ref{distri_cnt}) to
calculate the cluster concentrations are the formation energies
relative to the solid solution,
\begin{equation}
\Delta G_{n_{\textrm{X}}}(x^{0}_{\textrm{X}}) = G_{n_{\textrm{X}}}
- 2 n_{\textrm{X}} \mu(x^{0}_{\textrm{X}})
\label{direct_formation_eq},
\end{equation}
where $\mu(x^{0}_{\textrm{X}}) = \left(
\mu_{\textrm{X}}(x^{0}_{\textrm{X}}) -
\mu_{\textrm{Al}}(x^{0}_{\textrm{X}}) \right)/2$ is the effective
potential, \ie a Lagrange multiplier imposing that the nominal
concentration of the solid solution is equal to the concentration
of solute contained in the clusters as given by
Eq.~(\ref{conc_def}). As in the capillary approximation, this
formation energy can be divided into a volume and an interface
contribution:
\begin{equation}
\begin{split}
\Delta G_{n_{\textrm{X}}}(x^{0}_{\textrm{X}})  = &
4n_{\textrm{X}} \Delta G^{nuc}(x^{0}_{\textrm{X}}) \\
&+ \left(36\pi {n_{\textrm{X}}}^2\right)^{1/3} a^2
\sigma_{n_{\textrm{X}}} ,
\end{split}
\end{equation}
where we have defined the nucleation free energy
\begin{equation}
\begin{split}
\Delta G^{nuc}(x^{0}_{\textrm{X}}) =&
\left( 12\omega^{(1)} + 6 \epsilon_{\textrm{XX}}^{(1)} - 6 \epsilon_{\textrm{AlAl}}^{(1)} \right. \\
& \left. + 3 \epsilon_{\textrm{XX}}^{(2)} - 3
\epsilon_{\textrm{AlAl}}^{(2)} -2 \mu(x^{0}_{\textrm{X}}) \right)
/ 4
\end{split}
\label{direct_Gnuc_eq}
\end{equation}
and the interface free energy
\begin{equation}
\begin{split}
a^2\sigma_{n_{\textrm{X}}} =&
-kT\left(36\pi {n_{\textrm{X}}}^2\right)^{-1/3} \\
& \ln{\left( \sum_{\alpha} D_{n_{\textrm{X}},\alpha} \exp{ \left(
- \delta H_{n_{\textrm{X}},\alpha} /kT \right) } \right)}.
\label{direct_sigma_eq}
\end{split}
\end{equation}
All information concerning the solid solution, \ie its nominal
concentration, is contained in the nucleation free energy whereas
the interface free energy is an intrinsic property of clusters,
which was already the case with the capillary approximation. The
main difference is that now the interface free energy depends on
cluster size. Perini \etal \cite{PER84} show that this size
dependence can be taken into account in the capillary
approximation by adding terms to the series~(\ref{Gn_eq}) of the
formation energy reflecting \emph{line} and \emph{point}
contributions.

\begin{figure}
\includegraphics[width=0.9\linewidth]{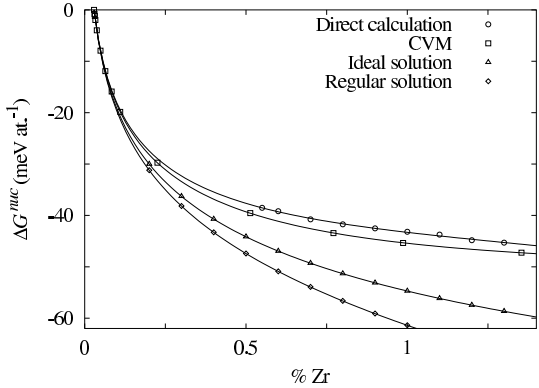}
\caption{Variation with the nominal concentration
$x^0_{\textrm{Zr}}$ of the nucleation free energy $\Delta G^{nuc}$
at $T=723$~K obtained with different approximation: direct
calculation of the cluster formation free energy
(Eq.~(\ref{direct_Gnuc_eq})) or capillary approximation with the
nucleation free energy given by the CVM calculation, the ideal
solid solution model, and the regular solid solution model.}
\label{direct_Gnuc_fig}
\end{figure}

We compare the nucleation free energy obtained from this direct
calculation of the cluster formation energies with the one that we
previously calculated with CVM in section~\ref{Gnuc_chap}
(Fig.~\ref{direct_Gnuc_fig}). The \emph{direct calculation} leads
to a slightly lower nucleation free energy in absolute value than
the CVM one. This mainly arises from the neglect of excluded
volume between the different clusters in the \emph{direct
calculation}. Nevertheless, the agreement is correct for all
temperatures and for both Al-Zr and Al-Sc systems. This shows that
these two approaches used to describe thermodynamics of the solid
solution, \ie the mean-field and the cluster descriptions, are
consistent.

\begin{figure}
\includegraphics[width=0.9\linewidth]{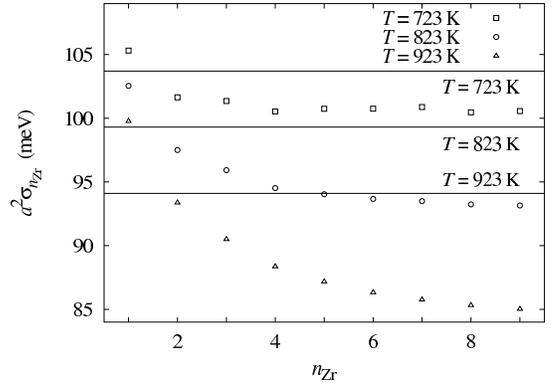}
\caption{Variation with the cluster size $n_{\textrm{Zr}}$ of the
interface free energy between the solid solution and Al$_3$Zr.
Symbols correspond to $\sigma_{n_{\textrm{Zr}}}$ as given by the
direct calculations of the cluster formation free energy
(Eq.~(\ref{direct_sigma_eq})) and lines to $\bar{\sigma}$, \ie the
Bragg-Williams calculation combined with the Wulff construction.}
\label{direct_sigma_fig}
\end{figure}

The interface free energy defined by Eq.~(\ref{direct_sigma_eq})
is decreasing with cluster size, the variation becoming more
important at higher temperatures (Fig.~\ref{direct_sigma_fig}).
The asymptotic limit is smaller than the interface free energy
$\bar{\sigma}$ that we calculated in section~\ref{sigma_chap}
using Wulff construction and Bragg-Williams approximation. This is
quite natural as Wulff construction predicts the cluster shape
costing least energy. We are thus missing some configurational
entropy by using it to compute an interface free energy
$\bar{\sigma}$ and we overestimate $\bar{\sigma}$. This error can
be neglected at low temperature ($T \leq 773$~K) where
precipitates show sharp interfaces but it increases with
temperature when precipitate shapes are becoming smoother.

\begin{figure}
\includegraphics[width=0.9\linewidth]{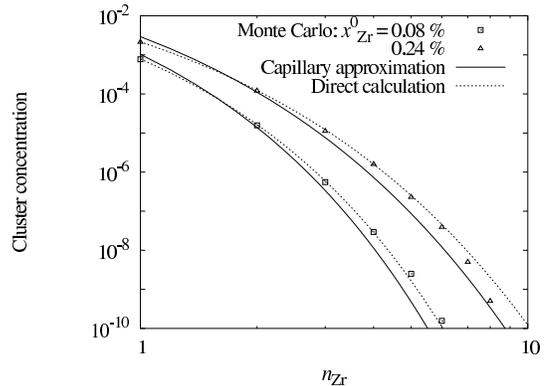}
\caption{Cluster size distribution of two aluminum solid solutions
of nominal concentrations $x^0_{\textrm{Zr}}=8\times10^{-4}$ and
$2.4\times10^{-3}$ at $T=873$~K. At this temperature, the
solubility limit is $x^{eq}_{\textrm{Zr}}=1.6\times10^{-3}$.
Symbols correspond to Monte Carlo simulations and lines to
prediction of classical nucleation theory as given by
Eq.~(\ref{distri_cnt}). To evaluate the cluster free energy of
formation, we use the capillary approximation (Eq.~(\ref{Gn_eq}))
with the nucleation free energy given by CVM for the continuous
line and the direct calculation (Eq.~(\ref{direct_partition_eq})
and (\ref{direct_formation_eq})) for the dashed line.}
\label{direct_distri_fig}
\end{figure}

We use this direct calculation of the cluster formation free
energy (Eq.~(\ref{direct_partition_eq}) and
(\ref{direct_formation_eq})) to predict cluster size distributions
in the solid solution and compare the results with the
distributions obtained with the capillary approximation
(Eq.~(\ref{Gn_eq})) combined to the CVM calculation. These two
models lead to similar distributions
(Fig.~\ref{direct_distri_fig}), indicating that the associated
thermodynamic descriptions are consistent. Nevertheless, the
distribution predicted by the direct calculation better reproduces
the ones measured during the Monte Carlo simulations. Thus, the
capillary model is good to describe thermodynamics of the solid
solution but it can be improved.

\begin{figure}
\includegraphics[width=0.9\linewidth]{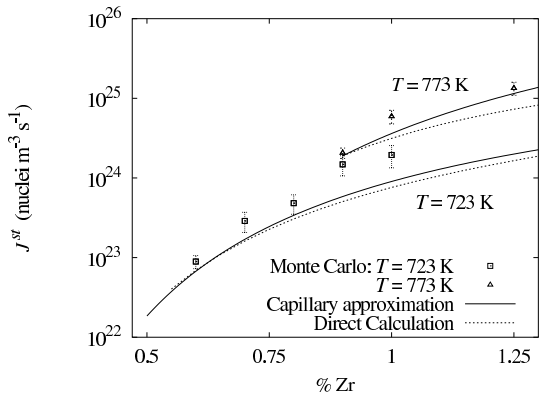}
\caption{Variation with nominal concentration $x^0_{\textrm{Zr}}$
and temperature of the steady-state nucleate rate $J^{st}$ for
Al$_3$Zr. Solid lines correspond to prediction of the classical
nucleation theory when using the capillary approximation with the
CVM calculation of the nucleation free energy and dashed lines
when using the direct calculation of the cluster formation free
energy. Symbols are measurements in Monte Carlo simulations. The
errorbars correspond to the uncertainty on the measurements of
$J^{st}$ due to the choice of the critical size corresponding to
each energetic model.} \label{Jstat_direct_fig}
\end{figure}

Comparing the steady-state nucleation rates predicted by the two
thermodynamic models with the ones measured in Monte Carlo
simulations (Fig.~\ref{Jstat_direct_fig}), we do not obtain any
improvement by using the direct calculation of cluster energy
instead of the capillary approximation. For low supersaturations,
both models are in reasonable agreement with Monte Carlo
simulations whereas for higher supersaturations discrepancies
appear. The direct calculation leads to a slightly lower
nucleation rate than the capillary approximation. This mainly
arises from a difference of the critical size: $n_{\textrm{X}}^*$
is usually 1 atom greater with the direct calculation than with
the capillary approximation. As the use of the direct calculation
improves the agreement for the cluster size distribution, the
discrepancy observed at high supersaturations is not due to a bad
description of the solid solution thermodynamics but may arise
from limitations of classical nucleation theory itself. The
assumption of a constant flux between the different size classes
made by this theory to solve the rate equations associated to the
cluster size evolution may not apply at high supersaturations.
This can be seen in our atomic simulations by the fact that, for
these supersaturations, the linear domain observed for the
variation with time of the number of precipitates and used to
define the steady-state nucleation rate is more restricted than in
the low supersaturation case shown on
Fig.~\ref{alsc_kinetic_picture}. One could try to improve the
agreement with atomic simulations by using more sophisticated
mesoscopic models like cluster dynamics
\cite{BIN76,MIR77,BIN87,GUY02} which do not need such a kinetic
assumption to solve the rate equations. Another improvement that
could be made to classical nucleation theory is to consider the
variation with the nominal concentration of the diffusion
coefficient of X atoms which would lead to a diffusion coefficient
different from the impurity one that we use.

\subsection{Other mean-field approximation}
\label{mean_field_chap}

Usually, one does not calculate the nucleation free energy with
CVM as we did in section \ref{Gnuc_chap} but one uses simpler
mean-field approximation to evaluate the chemical potentials
entering Eq.~(\ref{Gnuc}) of $\Delta G^{nuc}$. We test these other
approximations and see if they are reliable to be used with
classical nucleation theory.

The easiest approximation that can be used is the ideal solid
solution model in which one keeps only the configurational entropy
contribution in the expression of chemical potentials and
calculates this term within the Bragg-Williams approximation. This
leads to the following expression
\begin{eqnarray}
\Delta G^{nuc}_{\textrm{ideal}}(x^{0}_{\textrm{X}})  &=&
\frac{3}{4} \ln{ \left( \frac{1-
x^{eq}_{\textrm{X}}}{1-x^{0}_{\textrm{X}}} \right) }
\nonumber \\
&& + \frac{1}{4} \ln{ \left(
\frac{x^{eq}_{\textrm{X}}}{x^{0}_{\textrm{X}}} \right) } .
\label{Gnuc_ideal}
\end{eqnarray}

The exact expression of the nucleation free energy, \ie with the
enthalpic contribution, can be calculated within the
Bragg-Williams approximation too. This is called the regular solid
solution model and gives
\begin{eqnarray}
\Delta G^{nuc}_{\textrm{BW}}(x^{0}_{\textrm{X}}) &=& \Delta
G^{nuc}_{\textrm{ideal}}(x^{0}_{\textrm{X}}) +\Omega \left[
\frac{3}{4} \left( {x^{eq}_{\textrm{X}}}^2 -
{x^{0}_{\textrm{X}}}^2 \right)
\nonumber \right.\\
&& + \left. \frac{1}{4} \left( (1-x^{eq}_{\textrm{X}})^2 -
(1-x^{0}_{\textrm{X}})^2 \right) \right] . \label{Gnuc_BW}
\end{eqnarray}

Comparing all different mean-field approximations used to evaluate
the nucleation free energy (Fig.~\ref{direct_Gnuc_fig}), we see
that for low supersaturations all approximations are close, but
that for an increasing nominal concentration of the solid solution
discrepancies between the different approximations are becoming
more important. Both ideal and regular solid solution models
overestimate the nucleation free energy compared to the CVM and
the direct calculations. The discrepancy is even worse when all
contributions, \ie the enthalpic and entropic ones, are considered
in the Bragg-Williams approximation. Thus, when the
supersaturations is becoming too important, the Bragg-Williams
approximation seems too rough to give a reliable approximation of
the nucleation free energy.

\begin{figure}
\includegraphics[width=0.9\linewidth]{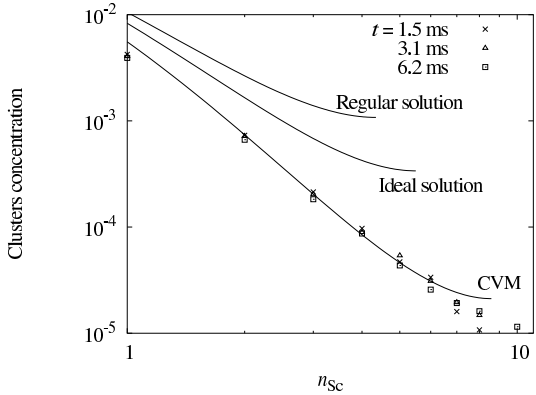}
\caption{Cluster size distribution of an aluminum solid solution
of nominal concentration $x^0_{\textrm{Sc}}=7.5\times10^{-3}$ at
$T=773$~K. Symbols correspond to Monte Carlo simulations and lines
to prediction of classical nucleation theory with the different
mean field approximations of the nucleation free energy.}
\label{mean_field_distri_fig}
\end{figure}

This becomes clear when combining these approximations of $\Delta
G^{nuc}$ with classical nucleation theory to predict cluster size
distributions. The ideal and the regular solid solution models
completely fail for high supersaturations to predict the cluster
size distributions observed during Monte Carlo simulations
(Fig.~\ref{mean_field_distri_fig}). The predicted critical size
$n^*_{\textrm{X}}$ is too small as it corresponds to a cluster
size in the observed stationary distribution and the predicted
probabilities for each cluster size are too high compared to the
observed ones. As the prediction of the steady-state nucleation
rate by classical nucleation theory is based on the predicted size
distribution, the ideal solution model and the Bragg-Williams
approximation lead to an overestimation of $J^{st}$ too. Thus the
use of CVM to calculate nucleation free energy really improves
agreement with atomic simulations compared to more conventional
mean-field approximations. This arises from the fact that order
effects are not taken into account in Bragg-Williams approximation
whereas they are in CVM. These order effects correspond to a
strong attraction for first nearest neighbors and a strong
repulsion for second nearest neighbors between Al and Zr atoms as
well as Al and Sc atoms. They are the reason at the atomic scale
why a supersaturated aluminum solid solution evolves to lead to
the precipitation of a L1$_2$ compound. Therefore one must fully
consider these order effects when modelling kinetics of
precipitation.

\section{Conclusions}

We built an atomic kinetic model for Al-Zr and Al-Sc binary
systems so as to be as close as possible to the real systems.
Thanks to this model, we were able to simulate at an atomic scale
kinetics of precipitation of the L1$_2$ ordered compounds Al$_3$Zr
and Al$_3$Sc.

From this atomic model we deduced the corresponding interface and
nucleation free energies which, with the diffusion coefficients,
are the only parameters required by mesoscopic models like
classical nucleation theory. When CVM is used to calculate the
nucleation free energy we showed that the capillary approximation
leads to a satisfying thermodynamic description of the solid
solution. If one wants to improve this description, one can
calculate directly formation free energies of the different size
clusters. This leads to a better description of the thermodynamic
behavior of the solid solution, as the agreement on cluster size
distribution is better, but it does not dramatically change
predictions of the classical nucleation theory. This shows that
the capillary approximation is reasonable. From the kinetic point
of view, classical nucleation theory assumes that evolution of the
different clusters is governed by the long range diffusion of
monomers. For Al-Zr and Al-Sc systems, it appears to be a good
assumption as we checked that di-, tri-, and 4-mers do not diffuse
and that the steady-state nucleation rates measured in Monte Carlo
simulations are in good agreement with predictions of the
classical nucleation theory. Discrepancies appear at higher
supersaturations which may be due to the dependence of the
diffusion coefficient with the solute concentration of the
metastable solid solution or to the limits of the classical
nucleation theory which requires the nucleation regime to be
separated from the growth regime. Nevertheless, the nucleation
model was built on purpose to predict kinetics at low
supersaturations for which kinetic Monte Carlo simulations are not
tractable.

On the other hand, when one uses less sophisticated mean field
approximation than CVM like the Bragg-Williams approximation to
calculate the nucleation free energy, predictions of the classical
nucleation theory completely disagrees with Monte Carlo
simulations, especially when supersaturations are too high. This
shows that short range order effects which are naturally
considered in CVM must be taken into account so as to build a
kinetic mesoscopic model based on a reasonable physical
description. This is expected to be the case for all systems where
order effects are important and thus for systems leading to the
precipitation of an ordered compound.

\begin{acknowledgments}
The authors are grateful to Dr J. Dalla Torre, Dr B. Legrand, Dr F. Soisson,
and Dr G. Martin for their invaluable help and advice
on many aspect of Monte Carlo simulations and classical nucleation theory.
They would like to thanks too Dr Y. Le Bouar and Dr A. Finel for helpful discussions
on interface free energy and low temperature expansions.
This work was funded by the joint research program ``Precipitation''
between Pechiney, Usinor, CNRS, and CEA.
\end{acknowledgments}

\appendix*
\section{Wulff construction}
\label{Wulff_chap}

We use the Wulff construction \cite{PORTER,CHRISTIAN} so as to
define an isotropic interface free energies $\bar{\sigma}$ from
the free energies $\sigma_{100}$, $\sigma_{110}$, and
$\sigma_{111}$. This construction allows us to determine
precipitate real shape and to associate with it $\bar{\sigma}$
which corresponds to the same interface energy for a spherical
precipitate having the same volume.

Al$_3$X precipitates will show facets in the [100], [110], and
[111] directions if the following conditions are met:
\begin{eqnarray*}
&\sqrt{2}/2\ \sigma_{100} < \sigma_{110} < \sqrt{2}\ \sigma_{100} ,\\
&\sqrt{6}/3\ \sigma_{110} < \sigma_{111} < 2\sqrt{6}/3\
\sigma_{110} - \sqrt{3}/3\ \sigma_{100}  .
\end{eqnarray*}
For Al$_3$Zr and Al$_3$Sc, with the set of parameters given by
table \ref{energies_tab}, this is true for all temperatures. Each
facet surface will then be proportional to
\begin{eqnarray}
\Gamma_{100} &=& 4( \sigma_{100} - \sqrt{2}\sigma_{110} )^2 \nonumber \\
&& -2( \sigma_{100} - 2\sqrt{2}\sigma_{110} + \sqrt{3}\sigma_{111} )^2  , \\
\Gamma_{110} &=& 2\sqrt{2}( -2\sigma_{100} + \sqrt{2}\sigma_{110} ) \nonumber \\
&&( \sqrt{2}\sigma_{110} - \sqrt{3}\sigma_{111} ) , \\
\Gamma_{111} &=& 3\sqrt{3}/2\ ( -\sigma_{100}^2 - 2\sigma_{110}^2
+ \sigma_{111}^2  ) \nonumber \\ && + 3/2\ \sigma_{100}(
4\sqrt{6}\sigma_{110} - 6\sigma_{111} )
 .
\end{eqnarray}
Considering a spherical precipitate with the same volume and the
same interface energy, one gets
\begin{equation}
\bar{\sigma} = \sqrt[3]{\frac{1}{4\pi} (6\sigma_{100}\Gamma_{100}
+ 12\sigma_{110}\Gamma_{110} + 8\sigma_{111}\Gamma_{111} )} .
\end{equation}


\end{document}